\documentclass[structabstract]{aa}  
\usepackage{graphicx}
\usepackage{txfonts}

\def\degr{\hbox{$^\circ$}}

\newcommand{\asecdot}[2]{\mbox{#1$\stackrel {\prime \prime}{_{\bf \cdot}}$#2}}

\begin{document}
   \title{High angular resolution imaging and infrared spectroscopy of CoRoT candidates  
\thanks{Based on observations obtained at the European Southern
  Observatory at Paranal, Chile in programmes 
  282.C-5015A, 282.C-5015B,  282.C-5015C, 
  285.C-5045A, and 285.C-5045B,
  086.C-0235A, 086.C-0235B, 
  088.C-0707A, 088.C-0707B,
  090.C-0251A,  090.C-0251B, and
  091.C-203(A)}
}
\authorrunning{Guenther et al.}
\titlerunning{HR-imaging and infrared spectroscopy of CoRoT-candidates}
\author{Guenther, E.W.\inst{1}, 
Fridlund, M.\inst{3,14}, 
Alonso, R.\inst{2,4,5}, 
Carpano, S.\inst{3}, 
Deeg, H. J.\inst{4,5},
Deleuil, M.\inst{6}, 
Dreizler, S. \inst{9},
Endl. M.\inst{7},
Gandolfi, D.\inst{3},
Gillon, M.\inst{8},
Guillot, T.\inst{10}, 
Jehin, E.\inst{8},
L\'eger, A.\inst{11}, 
Moutou, C.\inst{6}, 
Nortmann, L.\inst{9},  
Rouan, D.\inst{12},
Samuel, B. \inst{12}, 
Schneider, J.\inst{13},
Tingley , B. \inst{4,5,15}
          }
   \institute{Th\"uringer Landessternwarte Tautenburg, 07778 Tautenburg, Germany
              (\email{guenther@tls-tautenburg.de})
         \and
Observatoire de l'Universit\'e de Gen\`eve, 51 chemin des Maillettes, 1290
Sauverny, Switzerland
         \and
Research and Scientific Support Department, ESTEC/ESA, PO Box 299, 2200
AG Noordwijk, The Netherlands
         \and
Instituto de Astrof\'\i sica de Canarias, 38205 La Laguna, Tenerife, Spain
         \and
Dpto. de Astrof\'\i sica, Universidad de La Laguna, 38206 La Laguna,
Tenerife, Spain
         \and
Laboratoire d'Astrophysique de Marseille, 38 rue Fr\'ed\'eric  Joliot-Curie, 
13388 Marseille Cedex 13, France 
         \and
McDonald Observatory, The University of Texas at Austin, Austin, TX 78712, USA
         \and
University of Li\` ege, All\'ee du 6 ao\^ut 17, S. Tilman, Li\` ege 1, Belgium
         \and
Georg-August-Universit\"at, Institut f\"ur Astrophysik, 
Friedrich-Hund-Platz 1,  37077  G\"ottingen, Germany
         \and
Observatoire de la C\^ote d' Azur, Laboratoire Cassiop\'ee, BP 4229,
06304 Nice Cedex 4, France
         \and
Institut d'Astrophysique Spatiale, Universit\'e Paris-Sud 11, 91405 Orsay, France
         \and
LESIA, UMR 8109 CNRS, Observatoire de Paris, UVSQ, Universit\'e
Paris-Diderot, 5 place J. Janssen, 92195 Meudon, France
         \and
LUTH, Observatoire de Paris, CNRS, Universit\'e Paris Diderot; 5 place
Jules Janssen, 92195 Meudon, France
\and
Leiden Observatory, Leiden University, P.O.Box 9513, NL-2300 RA, Leiden, The Netherlands
\and
Department of Physics and Astronomy, Ny Munkegade 120
University of Aarhus, 8000, Aarhus C, Denmark
}
\date{Received Feb 3, 2013; accepted Feb 20, 2013}
  \abstract 
{Studies of transiting extrasolar planets are of key importance for
  understanding the nature of planets outside our solar system
  because their masses, diameters, and bulk densities can be measured.
  An important part of transit-search programmes is the removal of
  false-positives.  In the case of the CoRoT space mission, the majority of
  the false-positives are removed by a detailed analysis of the
  light curves and by seeing-limited imaging in- and
  out-of-transit. However, the critical question is how many of the
  candidates that passed all these tests are false positives.  Such
  false positives can be caused by eclipsing binaries, which are
  either related or unrelated to the targets.}
{For our study we selected 25 CoRoT candidates that have already been screened
  against false-positives using detailed analysis of the light curves
  and seeing-limited imaging, which has transits that are between 0.7
  and 0.05\% deep. Our aim is to search for companion-candidates that
  had not been recognized in previous observations.}
{We observed 20 candidates with the adaptive optics imager NaCo and 
18 with the high-resolution infrared spectrograph CRIRES.}
{We found previously unknown stars within 2\arcsec of the targets in
  seven of the candidates.  All of these are too faint and too close to the
  targets to have been previously  detected with seeing-limited
  telescopes in the optical.  Our study thus leads to the surprising
  results that if we remove all candidates excluded by the sophisticated analysis of the
  light-curve, as well as carrying out deep imaging with seeing-limited telescopes, still
  28-35\% of the remaining candidates are found to possess companions that are bright 
  enough to be false-positives.}
{Given that the companion-candidates cluster around the targets and
  that the J-K colours are consistent with physical companions,
  we conclude that the companion-candidates are more likely to be physical
  companions rather than unrelated field stars. }
\maketitle
%

\section{Introduction}

Studies of transiting extrasolar planets are of key importance for
understanding the nature of planets outside our Solar System, because
they allow the derivation of their masses, diameters, and hence their
bulk densities. While ground-based transit search programmes have made
interesting discoveries, the photometric accuracy limits them to
special cases. Space telescopes like CoRoT 
(COnvection ROtation and planetary Transits)
open up an entirely new
field of research as they permit the detection of very small planets
like CoRoT-7b (L{\'e}ger et al. \cite{leger09}).  While the detection
of small-sized planets is interesting by itself, what is really
required is the determination of the {\em radius} and {\em mass} of
the planets. The mass-density diagram is the most important diagnostic
to find out whether they are low-density gaseous planets like
Jupiter, "ocean planets" (L{\'e}ger et al. \cite{leger04}), or
high-density rocky planets like the Earth. The determination of the
mass of a low-mass planet is, however, very time-consuming. Such a huge
investment in observing-time can only be justified if it is very
likely that a transit is caused by an orbiting planet and not by
something else. Removing false-positives (FPs), i.e. physical
configurations mimicking a transit-like signal, is an essential part
of transit search programmes. As pointed out by Alonso et
al. (\cite{alonso04}), there are a number of tests for removing FPs.
Almenara et al.  (\cite{almenara09}) showed that 83\% of the initial
detections in the CoRoT fields IRa01, LRc01, and LRa01 are FPs that
can already be removed with a detailed analysis of the
CoRoT light curves (see also Brown et al. \cite{brown03}).  The
remaining 17\% of the candidates require additional observations.

CoRoT uses photometric masks generated by the on-board software for
measuring the brightness of the target stars. The exact size and form
of the masks depend on the brightness and the colour of the star as
well on as other constraints (Llebaria \& Guterman \cite{llebaria06}).
Given the size of the mask, which is typically of the order of
$35\arcsec\times23\arcsec$, it is not that
unusual that there are stars other than the target within it.
If these are eclipsing binaries and sufficiently bright, they could
also be FPs.  By taking an image during transit and one out of
transit, we can find out if such a star is an eclipsing binary or
not. Since these images are taken with seeing-limited telescopes, they
allow us to detect all potential FPs with distances larger than about
2\arcsec from the target. 
This means that seeing-limited imaging
allows $\geq 98\%$ of the FPs caused by field stars to be removed.  In
principle, the removal of FPs by the detailed analysis of the
light curves and the seeing-limited imaging should thus remove almost
all FPs.  This has been the subject of photometric follow programme of
CoRoT, which is described in more detail in Deeg et
al. (\cite{deeg09}).

The critical question thus is how many of the candidates that passed
all these tests are still going to be FPs.  Such
FPs can be caused by eclipsing binaries located within
2\arcsec of the targets. These could be either related or unrelated
to the targets.  Answering this question is 
interesting not only in
the context of the CoRoT survey but also in the context of other
similar surveys. Another
important aspect is that additional stars within the 
point spread function (PSF) of CoRoT
will change the depth of the transit. It is thus important to know
the contamination factor.
To answer these questions, we need to explore the area $<$
2\arcsec~from the target star. We thus obtained AO-imaging and
high-resolution spectroscopy in the 
 of 25 CoRoT candidates
that passed the screening using both the analysis of the CoRoT \,
light curves and imaging in- and out- of transit with seeing-limited
telescopes.  The candidates and the details of the screening against
FPs are described in Carpano et al (\cite{carpano09}; IRa01), in
Carone et al.  (\cite{carone12}; LRa01), in Cabrera et
al. (\cite{cabrera09}; LRc01), in Erikson (\cite{erikson12}, SRc01),
and in Cavarroc (\cite{cavarroc12}; SRa03 and LRa03).  

Although the seeing-limited imaging is not the subject of this article, we will
briefly describe the results obtained for the targets that we discuss.

\section{Observations}

\subsection{AO-imaging with NaCo}

Using the AO-facility instrument NaCo (Nasmyth Adaptive Optics System,
NAOS and Near-Infrared Imager and Spectrograph, CONICA) mounted on UT4
(Yepun), we obtained diffraction-limited images of 20 CoRoT targets in
the J-band. Table \ref{tab:names} gives an overview of the targets
that we observed.

As shown in Almenara et al. (\cite{almenara09}), diluted binaries are
the main source of FPs, particular for candidates with a
transit depth $\leq 0.5\%$.  Diluted binaries consist of a primary
star (A) and an eclipsing binary (B and C), which is usually much
fainter.  The three stars can form either a triple system or an
eclipsing binary that is in the fore- or background of the primary.

In the limiting cases, star C is too faint to be detected yet large
enough to occult B completely. This is the minimum brightness that an
FP can have (e.g. minimum brightness of star B).  The depth of the
transits detected and the minimum brightness of potential FPs in V and R
for the 20 targets observed with NaCo are given in Table
\ref{tab:NaCo}. The brightness of the FPs in V and R are calculated
from the depth of the transit and the brightness of the star. In the
case of $J_{FP}$, we calculate the brightness of potential FPs for the
case that the FP is a physical companion and for the case
that it is an unrelated background object.

We decided to observe the stars in the J-band with NaCo to
minimize the difference between the wavelength at which CoRoT observes
and the wavelength of the NaCo observations.  However, to
plan the NaCo observations, we have to know how deep the images have
to be so that all potential FPs can be detected. This means
that we have to know the typical colour index of the stars in the
field. The UCAC-2 catalog lists the brightness of the stars in the
$579$nm to $642$nm-band (label $V_{UCAC}$) and in the J-band (taken
from 2MASS; Skrutskie et al. \cite{2MASS}). Although CoRoT observes
the whole wavelength region from 370 to 950 nm, the instrument is most
sensitive in the wavelength range between 600 and 700 nm (Costes \&
Perruchot \cite{costes06}; Levacher \cite{levacher06}). The
wavelength range of the UCAC-2 catalogue thus is close to the
wavelength range of the peak sensitivity of CoRoT. Using this
catalogue, we derived the $V_{UCAC}-J$-colours of all stars within 20
arcmin from our targets. As expected, faint stars in the CoRoT fields
have red colours.  For the targets in LRa01, LRa02, LRa03, and LRa04 we
find $V_{UCAC}-J=1.6\pm1.0$, and for SRa01, SRa02, and SR03,
$V_{UCAC}-J=$$1.9\pm1.2$, $1.7\pm1.0$, and $2.0\pm0.9$,
respectively. For stars in LRc02 we derive $V_{UCAC}-J=2.9\pm1.4$.
Using these numbers, we estimate the minimum brightness of potential
FPs in column 5 in Table \ref{tab:NaCo}. We thus conclude that we have
to reach typically J=20 in order to be certain to detect all potential
FPs.

Using total on-source exposure times between 12.5 and 29.2 minutes,
our NaCo images reach a 3-$\sigma$ detection limits between $J=21.7$
and $J=22.4$. This is deep enough to detect FPs. Because we exposed
the individual images short enough so that the target stars are not
saturated, we can use them as photometric reference stars. In many
cases, the NaCo images also contain other stars that are listed in the
2MASS (Skrutskie et al. \cite{2MASS}).  We can thus determine the
photometric error by determining their brightness in NaCo images and
comparing these with the values given in 2MASS.

In three cases where we found companion-candidates (CCs), we
obtained J- and K-band images in order to constrain their nature.  The
detection limits are almost the same in both filters, although these
stars are brighter in the K-band and the Strehl ratio is also higher.

Six objects were observed in visitor mode in December 2010 and 12
objects in December 2011, the others were observed in
service mode. Although articles about CoRoT-7b and CoRoT-32b have or
are being published, we mention them in this article because they are
part of the same observing programme (L{\'e}ger \cite{leger09}; Gandolfi
et al. \cite{gandolfi2013}).  Except for CoRoT-7b, which was observed
with the S13 camera, all observations were taken with the S27-camera
which has an image scale of $0\farcs02715$ per pixel, and a
field of view of $\sim 27\arcsec$.  To detect faint background stars within 
the PSF of the primary
stars, we used the high-dynamic range mode of NaCo and adjusted the
individual exposure time so that they were not saturated.  We thus
used individual exposures (DITs) 
between 2 and 60 seconds, depending on the brightness of the
object.  To remove instrumental artifacts we rotated NaCo
typically nine times in position angle with steps of $10\degr$.

IRAF (Image Reduction and Analysis Facility)
routines were used to flat-field the data, remove cosmic rays
hits, remove detector artifacts, and derotate the individual
images so that north is up and east is left in all images. The final
images were then created by co-adding the individual images using a
kappa-sigma clipping algorithm after shifting them to the same
position and derotating them.  By combining images taken at different
rotation angles of the instrument, artifacts are very efficiently
removed because they rotate with the instrument. To search
for faint stars within the PSF of the targets, we constructed a
rotationally averaged PSF for each target, which we then subtracted
from it. This self-referencing avoids artifacts that are usually
introduced if a PSF of a standard star is subtracted because
standard stars never have exactly the same brightness and colour as
the target.  The self-referencing using a rotationally averaged PSF
works so well because the images also are created from frames that
are rotated before they are averaged.

In cases where we found stars within the PSF of the primary, we
measured the stellar brightness of the secondary after first
subtracting the PSF of the primary.  We did this subtraction in
several different ways to determine the photometric error
introduced by the process.

\subsection{NIR spectroscopy with CRIRES}

Although NaCo allows stars as close as \asecdot{0}{3} from
the target to be detected, it is still possible that there are stars within that
distance from the primary. The type of FP that is the most difficult to
exclude is a K- and/or M-type companion (Guenther \& Tal-Or
(\cite{guenther2010}). The best way to detect such
a companion is to obtain high-resolution NIR spectra. If a candidate
had a companion, we would detect lines that are specfic for a K-
and/or M-type companion star, like strong CO lines.  We thus obtained
high-resolution infrared spectra of 18 candidates with CRIRES (CRyogenic
high-resolution InfraRed Echelle Spectrograph) mounted on UT1 (Antu).

Because we are limited to stars bright enough to be used as natural
guide stars, we could not observe all candidates with CRIRES. As also
explained in Guenther \& Tal-Or (\cite{guenther2010}), the best
wavelength region is the K-band because the difference in brightness
between a G- and an M-star is much smaller in the K-band than at shorter
wavelengths. Using longer wavelengths is not useful, because the
sky brightness increases dramatically when going to the L or M-band.

We used two settings that are both well suited for detecting
late-type companions.  The first of these covered the wavelength range
2241.5 to 2281.4 nm (vacuum), which contains a number of prominent CaI
lines that are strong in K- and M-stars (Wallace \& Livingston
\cite{wallace92}). We used a slit width \asecdot{0}{3}, which gives us
a resolution of $\lambda / \Delta \lambda$ $\sim 60\ 000$.  The second
setting covered the wavelength region from 2284.1 to 2322.9 nm (in
vacuum), which contains a dense forest of CO-overtone lines.  These
lines are almost absent in F-stars but increase in strength from
spectral type G to mid-M.  Using a newly installed fixed slit with a
width of \asecdot{0}{4} gave a resolution of $\lambda / \Delta \lambda
\sim 48\ 000$.  An overview of the objects observed with CRIRES is
given in Table \ref{tab:CRIRES}.

The initial steps of the data reduction (removing artifacts,
flat-fielding, correcting for the non-linearity of the detector) were
done using the ESO pipeline and also independently with IRAF,
yielding similar results.  Since the spectra were taken by nodding the
star along the slit, we removed the sky background and the
bias offset by subtracting two spectra taken at different positions
along the slit. Each spectrum was then individually extracted and
wavelength calibrated using the telluric lines. The final spectra of
the stars were then created by averaging all individual spectra of
that star taken during the same observing night.  The telluric lines
were removed by using spectra taken of hot stars at the same airmass
also from the same night.  An example of a reduced spectrum is shown
in Fig.\,\ref{L2101CRIRES}.

\begin{table*}
  \caption{The objects}
\setlength{\tabcolsep}{3pt} 
 \begin{tabular}{ccllllll}
\noalign{\smallskip}
 \hline 
\noalign{\smallskip}
  CoRoT & CoRoT & RA     & DEC  & $V_{target}^1$  & $R_{target}^1$ & $J_{target}^2$ & 2MASS \\
  Win-ID  &             & h:m:s & d:m:s & [mag]               & [mag]                & [mag]           & name \\
\noalign{\smallskip}
 \hline  
\noalign{\smallskip}
LRa01$\_$E1$\_$0286  &         & $06^h 44^m 35\fs875$  & $+00\degr 00\arcmin 28\farcs440$       &  15.755 & $13.3\pm0.3$ & $11.06\pm0.03$ & 06443588+0000283 \\
LRa01$\_$E1$\_$2101  &          & $06^h 40^m 33\fs142$ & $+00\degr16\arcmin58\farcs944$         & $14.15\pm0.08$     & $13.51\pm0.01$ & $11.857\pm0.017$       & 06403313+0016590 \\
LRa01$\_$E1$\_$2240  &          & $06^h 43^m 37\fs337$ & $+00\degr16\arcmin51\farcs492$         & $15.22\pm0.03$     & $14.91\pm0.02$ & $13.806\pm0.030$      & 06433735+0016512 \\
LRa01$\_$E1$\_$4667  &         & $06^h 41^m 7\fs807$    & $+00\degr34\arcmin15\farcs096$         & 16.08                      & $15.3\pm0.3$ & $14.111\pm0.033$ & 06410780+0034152 \\
LRa01$\_$E1$\_$4719  &          & $06^h 43^m 42\fs427$ & $+00\degr49\arcmin47\farcs496$         & $15.88\pm0.04$     & $15.52\pm0.05$ & $14.399\pm0.044$     & 06434244+0049473 \\
LRa01$\_$E2$\_$0165  & CoRoT-7b & $06^h 43^m 49\fs454$ & $-01\degr03\arcmin46\farcs908$ & $12.93\pm0.04$     & $11.378\pm0.008$ & $9.773\pm0.024$ & 06434947-0103468 \\
LRa02$\_$E1$\_$1475  &         & $06^h 51^m  29\fs006$ & $-03\degr 49\arcmin 03\farcs468$        &  14.175                   & 13.4                      & $12.976\pm0.030$ & 06512900-0349034 \\
LRa02$\_$E1$\_$1715  &          & $06^h 51^m 18\fs046$ & $-03\degr22\arcmin15\farcs240$          & $14.84\pm0.10$     & $14.55\pm0.03$ & $13.525\pm0.021$ & 06511805-0322151 \\
LRa02$\_$E1$\_$4601  &          & $06^h 47^m 41\fs412$ & $-03\degr43\arcmin09\farcs469$ &                            & 15.1 & $13.596\pm0.021$ & 06474141-0343094 \\
LRa02$\_$E2$\_$1136  &          & $06^h 51^m 59\fs090$ & $-05\degr36\arcmin48\farcs888$          & $13.953\pm0.03$  & $13.68\pm0.04$ & $12.594\pm0.023$ & 06515909-0536488 \\
LRa02$\_$E2$\_$2057  &          & $06^h 50^m 50\fs266$ & $-05\degr00\arcmin35\farcs676$          & $14.889\pm0.04$  & $14.64\pm0.03$ & $13.731\pm0.026$ & 06505026-0500357 \\
LRa02$\_$E2$\_$3804  &          & $06^h 51^m 48\fs634$ & $-05\degr27\arcmin35\farcs496$          & $15.76\pm0.07$    & $15.47\pm0.06$ & $14.135\pm0.035$ & 06514863-0527354 \\
LRa03$\_$E2$\_$0678  &          & $06^h 09^m 33\fs156$ & $+04\degr41\arcmin12\farcs336$         & $13.55\pm0.03$    & $12.96\pm0.03$ & $11.391\pm0.026$ & 06093315+0441123 \\
LRa03$\_$E2$\_$0861  &          & $06^h 12^m 10\fs992$ & $+05\degr02\arcmin27\farcs132$         & $14.08\pm0.06$    & $13.67\pm0.06$ & $12.488\pm0.021$ & 06121099+0502270 \\
LRa03$\_$E2$\_$1326  &          & $06^h 13^m 50\fs765$ & $+05\degr18\arcmin08\farcs820$         & $14.51\pm0.04$    & $13.93\pm0.03$ & $11.910\pm0.021$ & 06135076+0518086 \\
LRa04$\_$E2$\_$0626  &          & $06^h 08^m 34\fs500$ & $+06\degr35\arcmin17\farcs030$         & 13.62                     & $13.50\pm0.01$ & $12.112\pm0.024$ & 06083449+0635171 \\
LRa06$\_$E2$\_$5287  &         & $06^h 45^m 13\fs771$  & $-00\degr 53\arcmin 26\farcs772$        & 15.76                     & $15.54\pm0.06$ & $13.791\pm0.030$ & 06451377-0053267 \\
LRa07$\_$E2$\_$3354  &         & $06^h 27^m  06\fs248$ & $+04\degr 32\arcmin 23\farcs924$       & 15.53 & $15.33\pm0.05$ & $13.86\pm0.022$  & 06270624+0432238 \\
LRc02$\_$E1$\_$0591  &         & $18^h 42^m 40\fs118$ & $+06\degr13\arcmin09\farcs300$ & $13.93\pm0.02$  & $13.56\pm0.02$ & $12.414\pm0.024$ & 18424010+0613088 \\
LRc07$\_$E2$\_$0158 & & $18^h 34^m 29\fs880$ & $+06\degr52\arcmin46\farcs533$                   & 12.7 & $12.18\pm0.03$           & $11.245\pm0.024$ & 18342987+0652466 \\ 
SRa01$\_$E1$\_$0770  &          &  $06^h 40^m 46\fs8$ & $+09\degr15\arcmin26\farcs8 $               & 13.9                              & 13.4 & $12.519\pm0.024$ & 06404684+0915267 \\
SRa02$\_$E1$\_$1011  &          & $06^h 29^m 30\fs157$ & $+06\degr16\arcmin30\farcs673$          &                                 & 13.6 & $12.571\pm0.023$ & 06293015+0616307 \\
SRa03$\_$E2$\_$2355  &          & $06^h 31^m 23\fs805$ & $+00\degr09\arcmin23\farcs630$          & 16.0                          & $15.27\pm0.09$ & $12.741\pm0.019$ & 06312379+0009239 \\
SRa03$\_$E2$\_$1073  &        & $06^h 29^m 48\fs583$  & $+00\degr 03\arcmin 51\farcs113$          &  14.6 & $14.5\pm0.4$ & $12.939\pm0.024$ & 06294859+0003512 \\
SRa04$\_$E2$\_$0106  & CoRoT-32b & $06^h 19^m 12\fs387$ & $-04\degr 38\arcmin 15\farcs382$  &  11.9 & 11.7 & $10.688\pm0.026$  & 06191238-0438154 \\
\noalign{\smallskip}
 \hline 
\noalign{\smallskip}
 \end{tabular}
 \label{tab:names}
\\
$^1$ EXODAT (Deleuil et al. \cite{deleuil09}), 
$^2$ 2MASS (Skrutskie et al. \cite{2MASS})\\
\end{table*}

\begin{table*}
  \caption{Summary of the results obtained with NaCo}
\setlength{\tabcolsep}{3pt} 
 \begin{tabular}{llccll}
\noalign{\smallskip}
 \hline  
\noalign{\smallskip}
  CoRoT & transit &  $V_{FP}^2$ &  $R_{FP}^2$ & $J_{FP}^3$ & results: \\
  Win-ID  & depth &   [mag] &  [mag] &  [mag] &  \\
\noalign{\smallskip}
 \hline  
\noalign{\smallskip}
LRa01$\_$E1$\_$0286   & 0.04\%      & 24.3 & 21.8 & $19.6-22.6$ & closest star at $5\farcs0$ distance\\ 
LRa01$\_$E1$\_$2101   & 0.09\%      & 21.8 & 21.1 & $19.5-20.2$ & companion-candidate: $J=16.3\pm0.1$, sep. $1\farcs8$ \\ 
LRa01$\_$E1$\_$2240   & 0.05\%      & 23.5 & 23.2 & $21.9-22.1$ & closest star at $5\farcs6$ distance  \\ 
LRa01$\_$E1$\_$4719   &  0.06\%     & 23.9 & 23.6 & $22.3-22.5$ & companion: G9V, $J=15.8\pm0.1$, sep. $0\farcs8$ \\ 
LRa01$\_$E2$\_$0165$^1$ & 0.05\% & 21.2 & 19.6 & $18.0-19.6$ & no star with $5 \arcsec$, CoRoT-7b \\ 
LRa02$\_$E1$\_$1715   & 0.02\%      & 24.1 & 23.8 & $22.5-22.8$ & companion: M4V, $J=18.8\pm0.1$, sep. $1\farcs5$ \\ 
LRa02$\_$E1$\_$4601  & 0.3\%  &         & 21.4 & $19.8-19.9$ & no companion candidate found  \\ 
LRa02$\_$E2$\_$1136   &  0.3\%        & 20.3 & 20.0 & $18.7-18.9$ & companion: K4V-K5V, $J=14.5\pm0.1$, sep. $0\farcs4$ \\ 
LRa02$\_$E2$\_$2057   & 0.07\% & 22.8 &  22.5 & $21.2-21.6$ & closest star at $5\farcs1$ distance \\ 
LRa02$\_$E2$\_$3804   &  1.0\% & 20.8 & 20.5 & $19.1-19.2$ & closest star at $10\farcs0$  distance \\ 
LRa03$\_$E2$\_$0678   & 0.1\%  & 21.1 & 20.5 & $18.9-19.5$ &  closest star at $9\farcs8$ distance \\ 
LRa03$\_$E2$\_$0861   & 0.1\% & 21.6 & 21.2 & $19.9-20.0$ &  companion-candidate: $J=16.4\pm0.1$, sep. $1\farcs1$ \\ 
LRa03$\_$E2$\_$1326   & 0.7\% & 19.9 & 19.3 & $17.3-18.3$ & closest star at $8\farcs3$ distance \\ 
LRa04$\_$E2$\_$0626   & 0.2\% & 20.4 & 20.2 & $18.7-18.8$ & companion-candidate: $J=16.8\pm0.1$,  sep. $0\farcs9$ \\ 
LRa06$\_$E2$\_$5287    & 0.2\% & 22.5 & 22.3 & $20.1-20.9$ & closest star with $J=20.3\pm0.2$ at $3\farcs6$ distance \\ 
LRc02$\_$E1$\_$0591 & 0.2\% & 20.7 & 20.3 & $17.8-19.2$ & see discussion in the text.\\ 
LRc07$\_$E2$\_$0158         & 0.03\% &  21.5 &  21.0 & $18.6-20.1 $  & companion-candidate: $J=14.6\pm0.1$,  sep. $0\farcs9$ \\ 
SRa01$\_$E1$\_$0770  & 0.3\%  & 22.7 & 22.2 & $20.8-21.3$ & closest star at $6\farcs2$ distance \\ 
SRa02$\_$E1$\_$1011   &  0.1\% & - & 21.1 & $ 19.9-20.0$ & closest star at $8\farcs9$ distance \\ 
SRa03$\_$E2$\_$2355   & 0.6\% & 21.6& 20.8 & $18.3-19.6$ & closest star at $3\farcs0$ distance  \\ 
\noalign{\smallskip}
 \hline 
\noalign{\smallskip}
 \end{tabular}
 \label{tab:NaCo}
\\
$^1$ CoRoT-7 \\
$^2$ A star causing a false-positive has to be brighter than this value.  \\
$^3$ Same as $^2$ but for the J-band. This value is derived from the
$<V-J>$-colours of stars in the field. \\
\end{table*}


\section{Stars with companion-candidates}
\label{Sec:StarsCC}

In this section we discuss the objects where we found faint stars
within 2\arcsec of our targets. In the following
we use the CoRoT Win-IDs for the targets because they are easier to
remember. For completeness, we list the Win-IDs, the 2MASS numbers,
and the position of all targets observed in Table \ref{tab:names}.

\begin{figure}
\includegraphics[height=.23\textheight,angle=0]{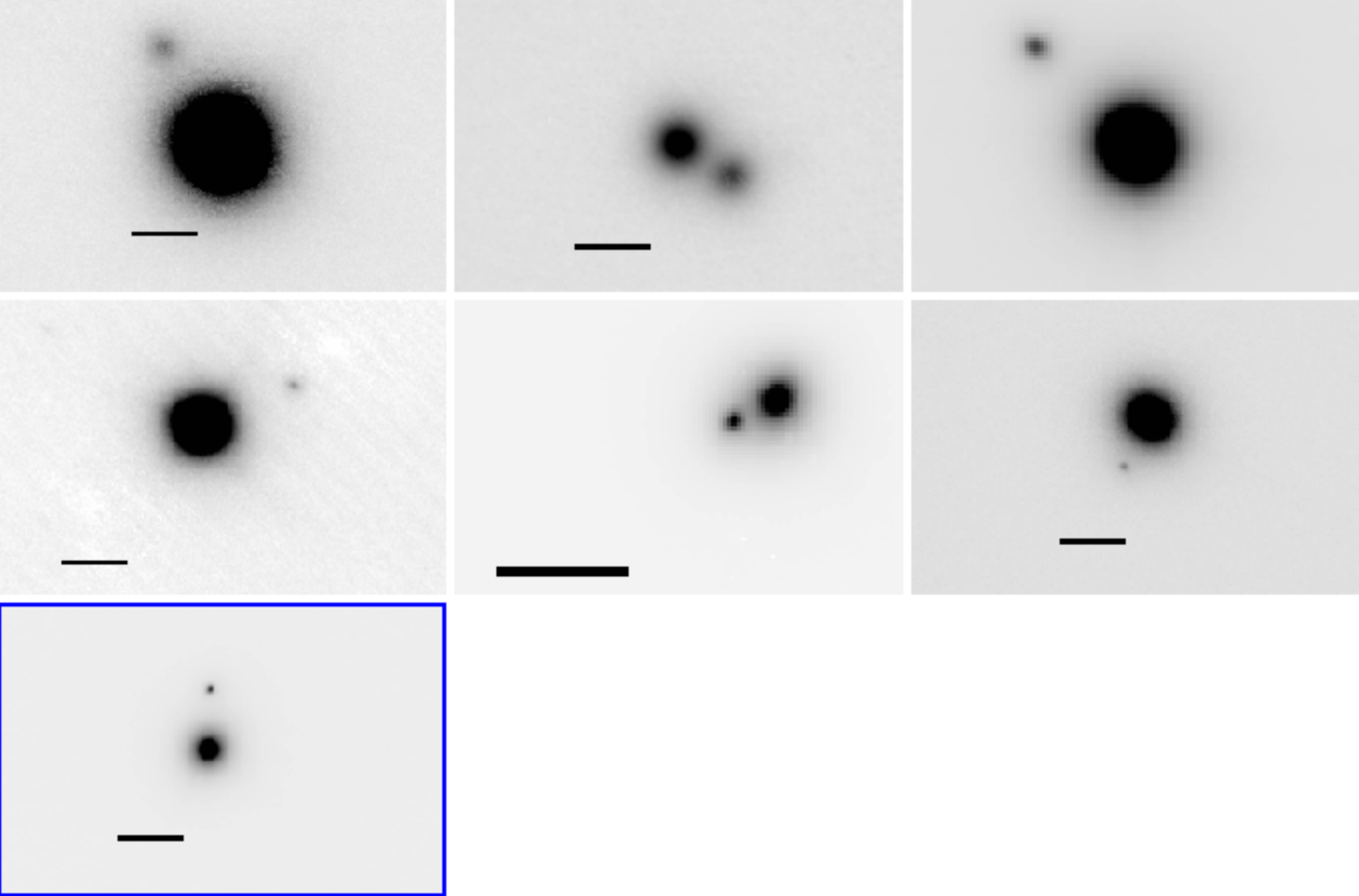}
\caption{NaCo images of objects with CCs were found.  Top row from
left to right: LRa01$\_$E1$\_$2101, LRa01$\_$E1$\_$4719, LRa03$\_$E2$\_$0861.  Middle
row from left to right: LRa02$\_$E1$\_$1715, LRa02$\_$E2$\_$1136, LRa04$\_$E2$\_$0626.
Bottom row: LRc07$\_$E2$\_$0158. North is up and east is left in all images. 
The dark line corresponds to one arcsec. Deatils about the object
are given  in Table \ref{tab:NaCo}.  }
\label{NaCoImages}
\end{figure} 

The results of the NaCo and CRIRES observations are summarized in
Table \ref{tab:NaCo} and \ref{tab:CRIRES}.  The objects where we found
CCs are LRa01$\_$E1$\_$2101, LRa01$\_$E1$\_$4719, LRa03$\_$E2$\_$0861, 
LRa02$\_$E1$\_$1715, LRa02$\_$E2$\_$1136, LRa04$\_$E2$\_$0626, and
LRc07$\_$E2$\_$0158. Fig.\,\ref{NaCoImages} shows images of the objects with
CCs. The dark line in this figure corresponds to one arcsec. Because
we observed the CCs in the infrared, we do not know how bright they
are in the optical. However, if they are unrelated to the targets, we
can give an estimate based on the typical colour index of field stars
in the vicinity of the targets.  If CCs are physical companions, we
can calculate their brightness from the flux in the infrared and the
spectral type and brightness of the primary in the optical and in the
infrared.  The NaCo results are summarized in Table \ref{tab:NaCo},

In total, 11 of the 20 stars observed with NaCo have a transit depth
$\leq 0.1\%$, and 9 have deeper ones.  We found CCs in three of the
objects with deep transits ($33\pm20\%$) and in five of targets with
shallow transits ($45 \pm20\%$). Although this is still a small number
statistics, the result is not surpring because fainter stars could
potentially be source of FPs for a shallower transit.

\begin{figure}
\includegraphics[height=.25\textheight,angle=0]{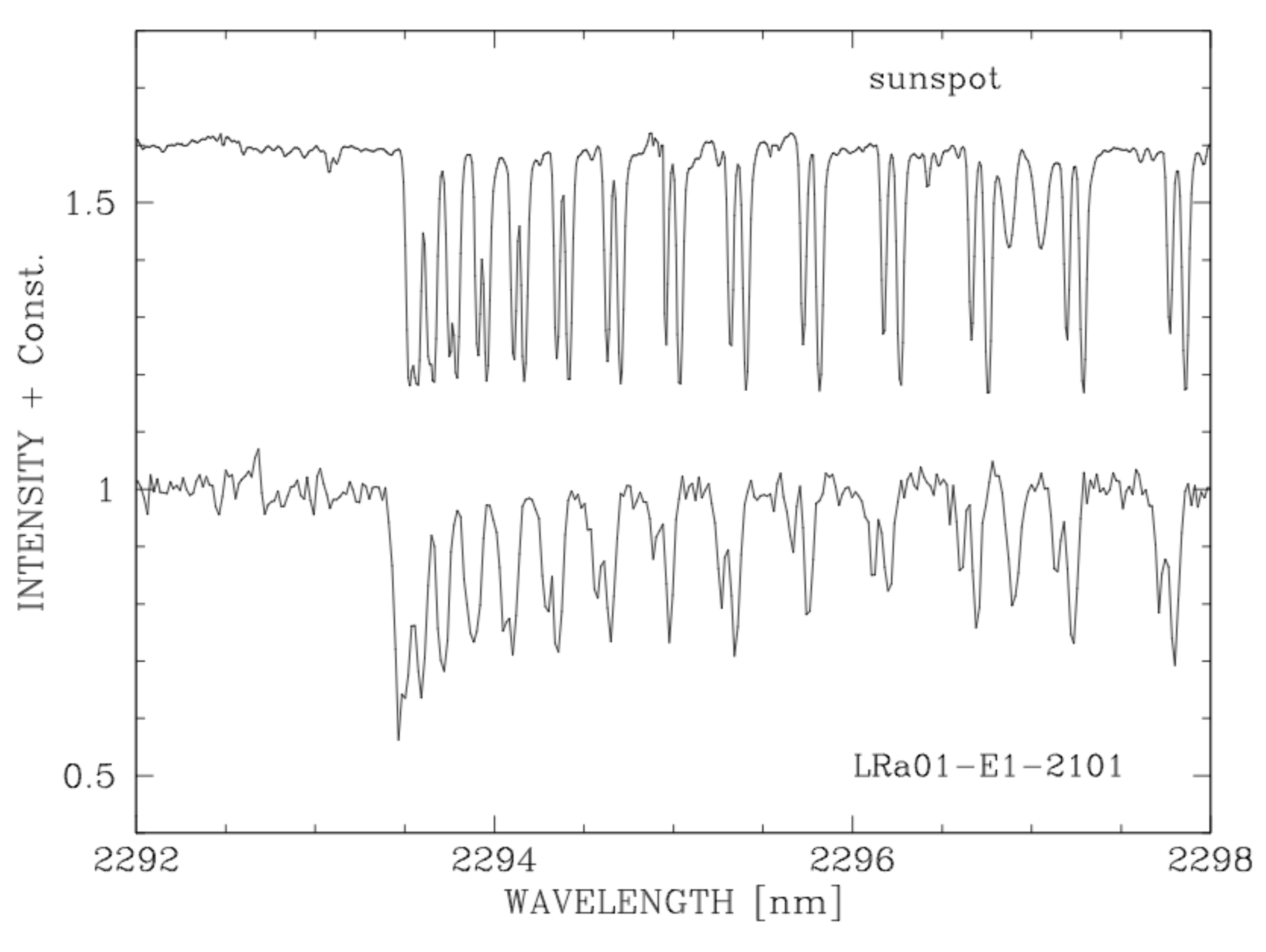}
\caption{Part of the CRIRES spectrum of LRa01$\_$E1$\_$2101 and a spectrum
of a sunspot for comparison. The CO lines are seen in both spectra.}
\label{L2101CRIRES}
\end{figure}

\begin{figure}
\includegraphics[height=.25\textheight,angle=0]{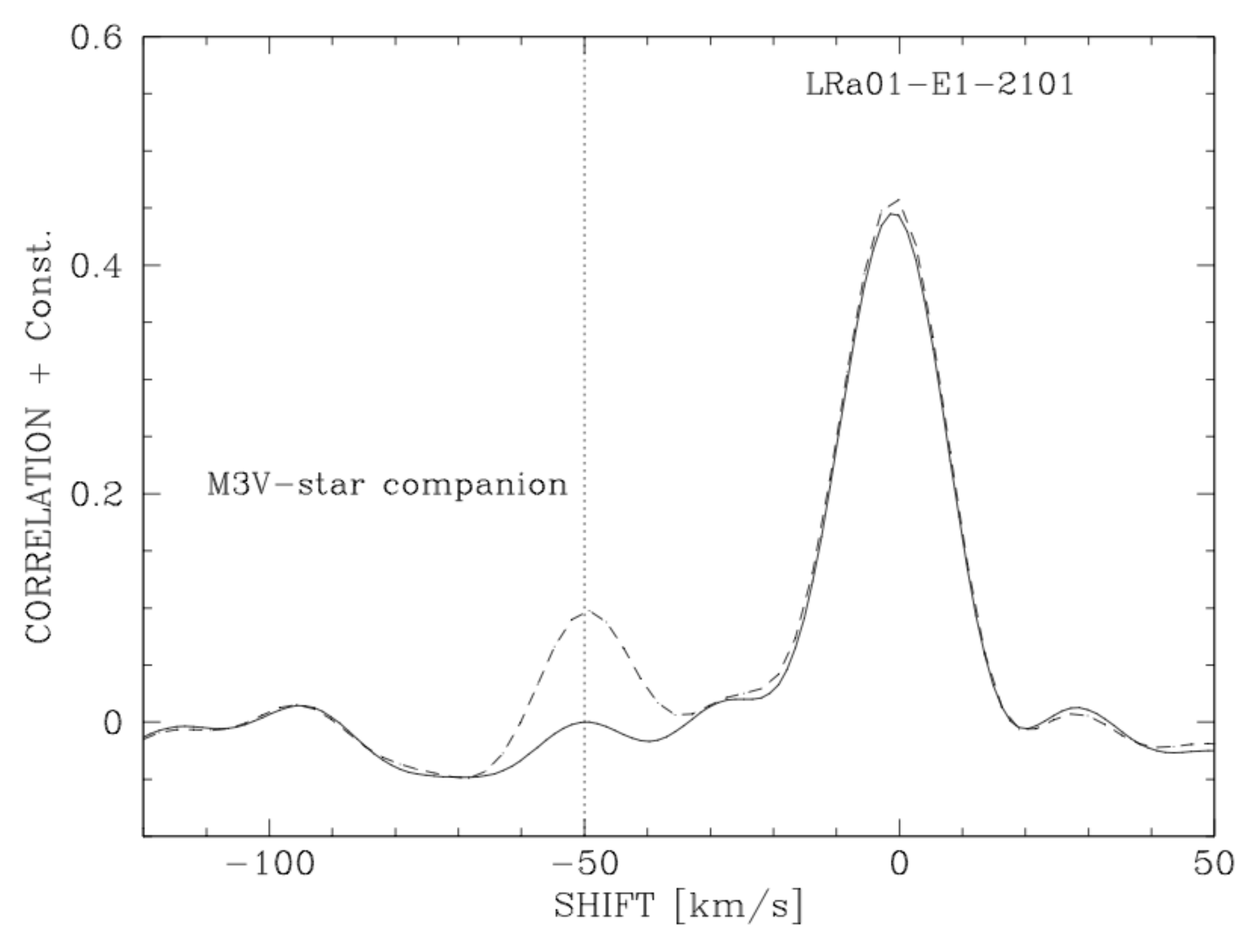}
\caption{Cross-correlation function of LRa01$\_$E1$\_$2101 (full line)
  together with a simulated cross-correlation function of
  LRa01$\_$E1$\_$2101 with a hypothetical M3V star added (dashed line).
  An M3V companion can thus be excluded.}
\label{L2101cc}
\end{figure}

The estimated brightness of the CCs in the optical is given in Table
\ref{tab:CCoptical}. The objects with CCs are discussed individually
in Appendix~\ref{Sec:StarsCCdetailed}. As an example for the CRIRES
spectra that we have taken, we show in Fig.\,\ref{L2101CRIRES} a
section of the spectrum containing the CO lines of LRa01$\_$E1$\_$2101. As
an example of how we exclude companion stars using the
cross-correlation technique, we show in Fig.\,\ref{L2101cc} a
simulated cross-correlation function of LRa01$\_$E1$\_$2101 with a
hypothetical M3V star added.
 
\begin{figure}
\includegraphics[height=.25\textheight,angle=0]{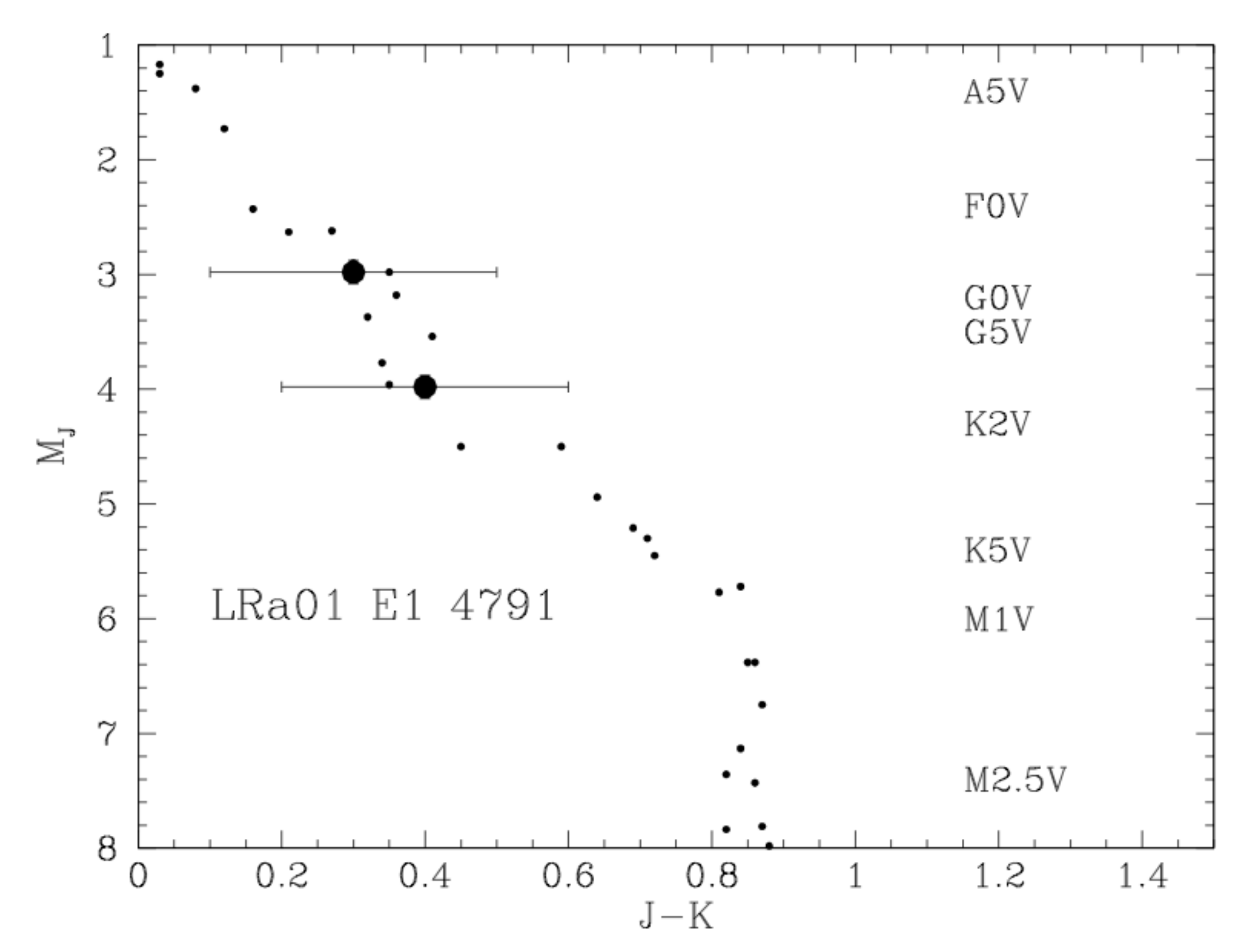}
\caption{Colour-magnitude diagram of the two stars of LRa$\_$E1$\_$4719
  (big dots). For comparison, we also show the brightness and J-K
  colour of standard stars (small dots) as given in L{\'e}pine et
  al. (\cite{lepine09}), Henry et al. (\cite{henry2006}), and Bilir et
  al. (\cite{bilir09}).}
\label{4719phot}
\end{figure}

\begin{figure}
\includegraphics[height=.25\textheight,angle=0]{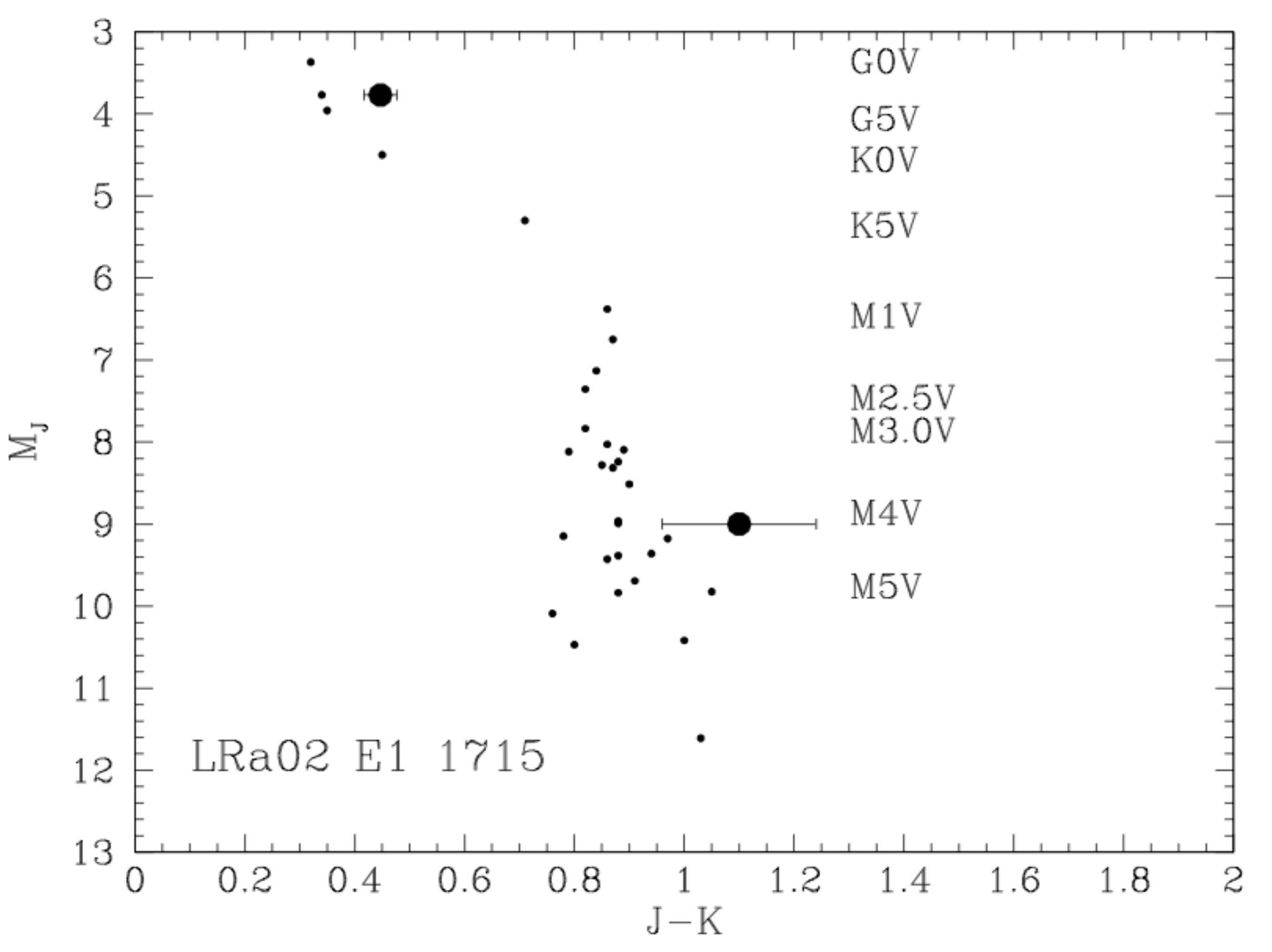}
\caption{Same as Fig.\,\ref{4719phot}  but for 
LRa$\_$E1$\_$1715.}
\label{1715phot}
\end{figure}

\begin{figure}
\includegraphics[height=.25\textheight,angle=0]{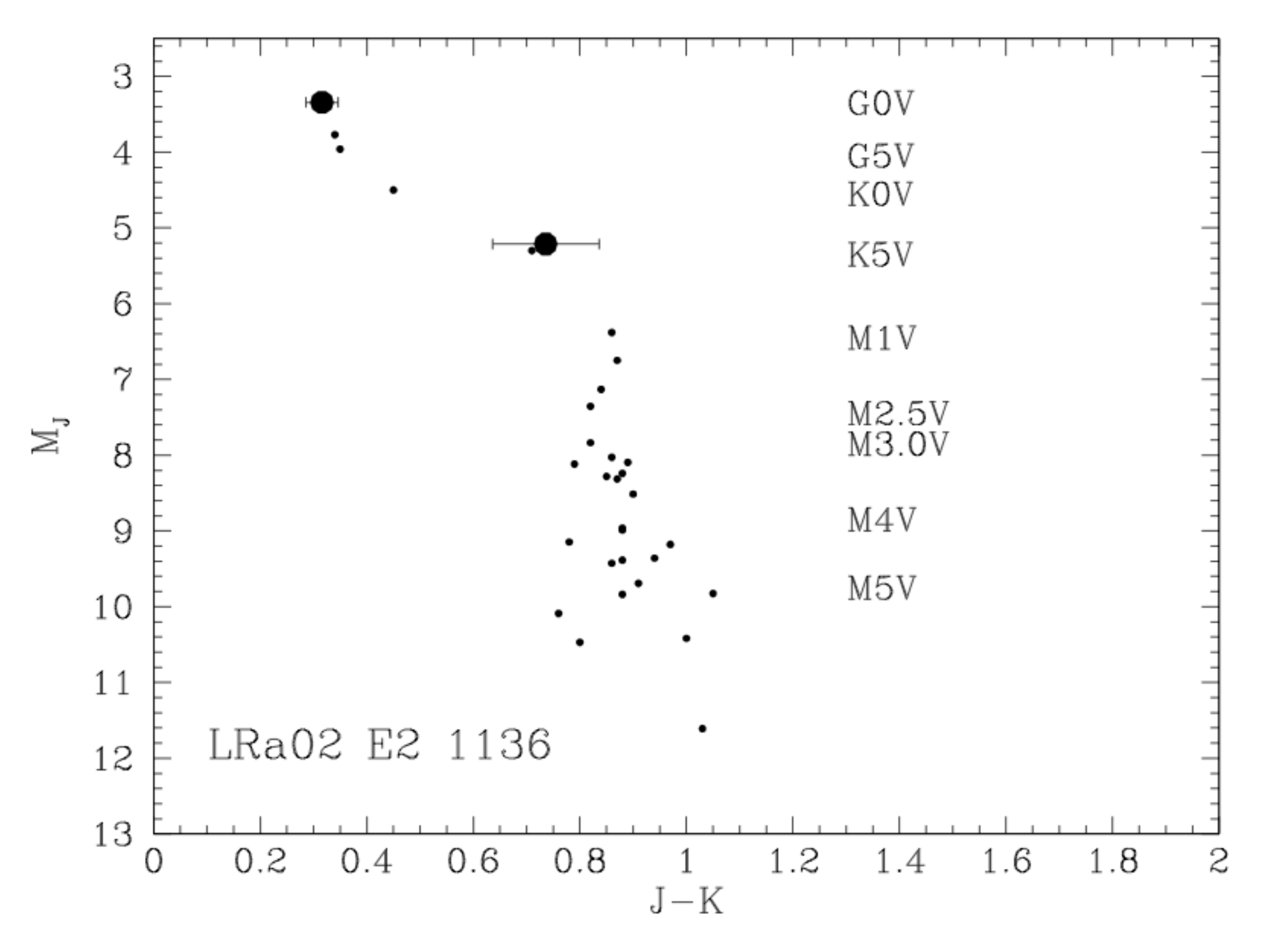}
\caption{Colour-magnitude diagram of the two stars of LRa02$\_$E2$\_$1136.}
 \label{1136phot}
\end{figure}

We obtained J- and K-band images
for LRa$\_$E1$\_$4719, LRa$\_$E1$\_$1715, and LRa02$\_$E2$\_$1136. 
Figs. ~\ref{4719phot}, \ref{1715phot}, and
\ref{1136phot} show the colour-magnitude diagram of the stars. The
absolute brightnesses of the CCs are calculated assuming that they are
at the same distance as the targets. The J-K colours are derived from
the observations. For comparison, we also show the brightness and J-K
colour of standard stars (small dots) as given in L{\'e}pine et
al. (\cite{lepine09}), Henry et al. (\cite{henry2006}), and Bilir et
al. (\cite{bilir09}). In all cases the J-K colours are consistent with
physical companions. Whether it is more likely that the CCs are
physical companions or unrelated backgrounds stars will be discussed
in Section~\ref{Sec:Discussion}.

\begin{table}
  \caption{Summary of the results obtained with CRIRES}
\setlength{\tabcolsep}{3pt} 
 \begin{tabular}{llccl}
\noalign{\smallskip}
 \hline 
\noalign{\smallskip}
  CoRoT & spec- &     & wavelength & spec- \\
               & type  & $K_{target}^2$ & [nm] & type \\
  Win-ID  & target & [mag] & (vac)    & comp.$^5$ \\
\noalign{\smallskip}
 \hline  
\noalign{\smallskip} 
LRa01$\_$E1$\_$0286  & G0V$^3$       & $10.319\pm0.024$  & 2276.9-2325.5 & see text \\
LRa01$\_$E1$\_$2101  & K6V         & $11.165\pm0.023$  & 2284.1-2322.9 & $\leq$M3.5V \\
LRa01$\_$E1$\_$4667  &  K2V             & $13.497\pm0.049$  & 2276.9-2325.5 & $\leq$M3.5V \\
LRa01$\_$E2$\_$0165$^2$ & G9V         & $8.734\pm0.022$ & 2284.1-2322.9 & $\leq$M5V \\
LRa02$\_$E1$\_$1475  &  A4V             & $12.676\pm0.037$  & 2276.9-2325.5 & $\leq$F6V \\
LRa02$\_$E1$\_$1715  & G2V$^6$  & $13.078\pm0.035$ & 2241.5-2281.4 & $\leq$M0V \\
LRa02$\_$E1$\_$4601 & K1V          & $12.924\pm0.023$ & 2241.5-2281.4 & $\leq$M1V \\
LRa02$\_$E1$\_$4601 & K1V          & $12.924\pm0.023$ & 2284.1-2322.9 & $\leq$M2.5V \\
LRa02$\_$E2$\_$1136 & G0V          & $12.169\pm0.024$ & 2241.5-2281.4 & $\leq$M0V \\
LRa02$\_$E2$\_$2057  & F8V$^{4,6}$ & $13.438\pm0.041$ & 2241.5-2281.4 & $\leq$M0V \\
LRa03$\_$E2$\_$0678  & K5V         & $10.706\pm0.019$ & 2241.5-2281.4 & $\leq$M1V \\
LRa03$\_$E2$\_$0861  & G8V$^4$ & $11.981\pm0.023$ & 2241.5-2281.4 & $\leq$K5V \\
LRa04$\_$E2$\_$0626  & F9V         & $11.692\pm0.021$ & 2284.1-2322.9 & $\leq$ M1V \\
LRa06$\_$E2$\_$5287  & G0V$^3$      & $13.193\pm0.026$  & 2276.9-2325.5 & see text \\
LRa07$\_$E2$\_$3354  & B9V$^3$      & $13.433\pm0.043$  & 2276.9-2325.5 & A6V \\ 
SRa01$\_$E1$\_$0770  & F9V         & $12.193\pm0.027$ & 2284.1-2322.9 & $\leq$ M0V \\
SRa02$\_$E1$\_$1011  & F6V$^3$ & $11.988\pm0.023$ & 2284.1-2322.9 & $\leq$ M0V \\
SRa03$\_$E2$\_$1073  & F3V$^3$      & $12.462\pm0.026$  & 2276.9-2325.5 & M0V \\
SRa04$\_$E2$\_$0106  & F5IV         & $10.413\pm0.023$  & 2276.9-2325.5 & M3V \\
\noalign{\smallskip}
 \hline 
\noalign{\smallskip}
 \end{tabular}
 \label{tab:CRIRES}
\\
$^1$ Brightness of target taken from 2MASS (Skrutskie et al. \cite{2MASS}) \\
$^2$ CoRoT-7, $^3$ TLS-NASMYTH spectrograph,  $^4$ EXODAT\\
$^5$ Latest spectral type of a hypothetical companion that can be excluded. \\ 
$^6$ Luminosity class IV not excluded. \\ 
\end{table}

\section{Stars without candidate companions}
\label{Sec:StarsNoCC}

The 18 objects were we did not find any CCs within 2 \arcsec of our
targets are discussed in Appendix~\ref{Sec:StarsNoCCdetailed}.  Eight
of these were observed with NaCo and CRIRES. These are: LRa01$\_$E1$\_$0286,
LRa01$\_$E2$\_$0165 (CoRoT-7), LRa02$\_$E1$\_$4601, LRa02$\_$E2$\_$2057,
LRa03$\_$E2$\_$0678, LRa06$\_$E2$\_$5287, SRa01$\_$E1$\_$0770 and SRa02$\_$E1$\_$1011.  Five
objects were only observed with NaCo. These are: LRa01$\_$E1$\_$2240,
LRa02$\_$E2$\_$3804, LRa03$\_$E2$\_$1326, LRc02$\_$E1$\_$0591, and
SRa03$\_$E2$\_$2355.  For these objects the NaCo images alone do not allow
to fully exclude faint companion star with a separtion of less than
$0\farcs8$.  Another five were only observed with CRIRES.  These are:
LRa01$\_$E1$\_$4667, LRa02$\_$E1$\_$1475, LRa07$\_$E2$\_$3354, SRa03$\_$E2$\_$1073,
SRa04$\_$E2$\_$0106 (CoRoT-32).  Given that CRIRES is an AO-instument, we
used the aquisition images taken in the K-band in order to exclude
companions with separtions larger than about $0\farcs8$. However, as
discussed in Appendix~\ref{Sec:StarsNoCCdetailed} these images are not
as deep as the NaCo-images.

\section{Discussion and conclusions}
\label{Sec:Discussion}

\subsection{Would it be possible to detect the companion-candidates with 
seeing-limited telescopes?}

We have sudied 25 CoRoT candidates. Of these, 13 were observed with
NaCo and CRIRES. CCs were found in seven of them. 
All of them were found in the NaCo images. Another seven objects were only observed
with NaCo, and another five only with CRIRES.  In two of the targets
observed with CRIRES, we detected very weak CO lines. However, since both are
G0V stars, it is possible that these are the weak CO lines from the
star itself. Depending on whether we should count only the objects
that have been observed with NaCo or all objects, we find that
the rate of targets with CCs is 28 or 35\%, respectively.

The discovery of so many CCs raises the question if it would have been
possible to detect them with seeing-limited telescopes. The properties
of the CCs found are given in Table \ref{tab:CCoptical}. The
candidates found either have a separation $\leq 1\arcsec$ or are 4 to
9 mag fainter in the optical regime and have a separation $\leq
2\arcsec$. Detecting such objects with a seeing-limited telescope is
not possible. It is thus not surprising that we did not detect these
stars before.

\begin{table*}
  \caption{Properties of the companion-candidates}
\setlength{\tabcolsep}{3pt} 
 \begin{tabular}{lcclccccc}
\noalign{\smallskip}
 \hline 
\noalign{\smallskip}
object & $J_{CC}^1$ & $K_{CC}^1$ & SpecType$^2$ & sep. & \multicolumn{2}{c}{pysical companion$^3$} &  \multicolumn{2}{c}{unrelated star$^4$}  \\
 & & & & & $V_{CC}$ & $V_{CC}-V_{primary}$  & $V_{CC, UCAC}$  & $V_{CC, UCAC}-V_{primary, UCAC}$\\
\noalign{\smallskip}
 \hline  
\noalign{\smallskip} 
LRa01$\_$E1$\_$2101CC  & $16.3\pm0.1$ &                        &                & $1\farcs8$ & $21.5$ & $8.6$ & 17.9 & 4.2 \\ 
LRa01$\_$E1$\_$4719CC  & $15.8\pm0.1$ & $15.4\pm0.2$ & G9V        & $0\farcs8$ & $17.6$ & $1.6$ & 17.5 & 1.2 \\ 
LRa02$\_$E1$\_$1715CC  & $18.8\pm0.1$ & $17.7\pm0.1$ & M4V        & $1\farcs5$ & $23.2$ & $8.4$ & 20.4 & 5.8 \\ 
LRa02$\_$E2$\_$1136CC  & $14.5\pm0.1$ & $13.6\pm0.1$ & K4V-K5V & $0\farcs4$ & $16.5$ & $2.6$ & 18.1 & 4.3 \\  
LRa03$\_$E2$\_$0861CC  & $16.4\pm0.1$ &                       &                 & $1\farcs1$ & $20.2$ & $6.1$ & 18.0 & 4.1 \\ 
LRa04$\_$E2$\_$0626CC  & $16.8\pm0.1$ &                       &                 & $0\farcs9$ & $21.2$ & $7.6$ & 18.4 & 5.0 \\ 
LRc07$\_$E2$\_$0158CC  & $14.6\pm0.1$ &                       &                & $0\farcs9$ & 17.8 & 5.1 &  16.2 & 3.5  \\ 
\noalign{\smallskip}
 \hline 
\noalign{\smallskip}
 \end{tabular}
 \label{tab:CCoptical}
\\
$^1$ Measured brightness of the CC in the J- and K-band.\\
$^2$ Spectral type of the CC as derived from the J-K colours. \\
$^3$ Calculated brightness of the CC and brightness difference between primary in the V-band, assuming it is a companion. \\
$^4$ Calculated brightness of the CC and brightness difference between primary in the $V_{UCAC}$ (579-642nm)
 band, assuming it is unrelated. \\
\end{table*}

\subsection{What is the nature of the companion-candidates?}

\begin{figure}
\includegraphics[height=.25\textheight,angle=0]{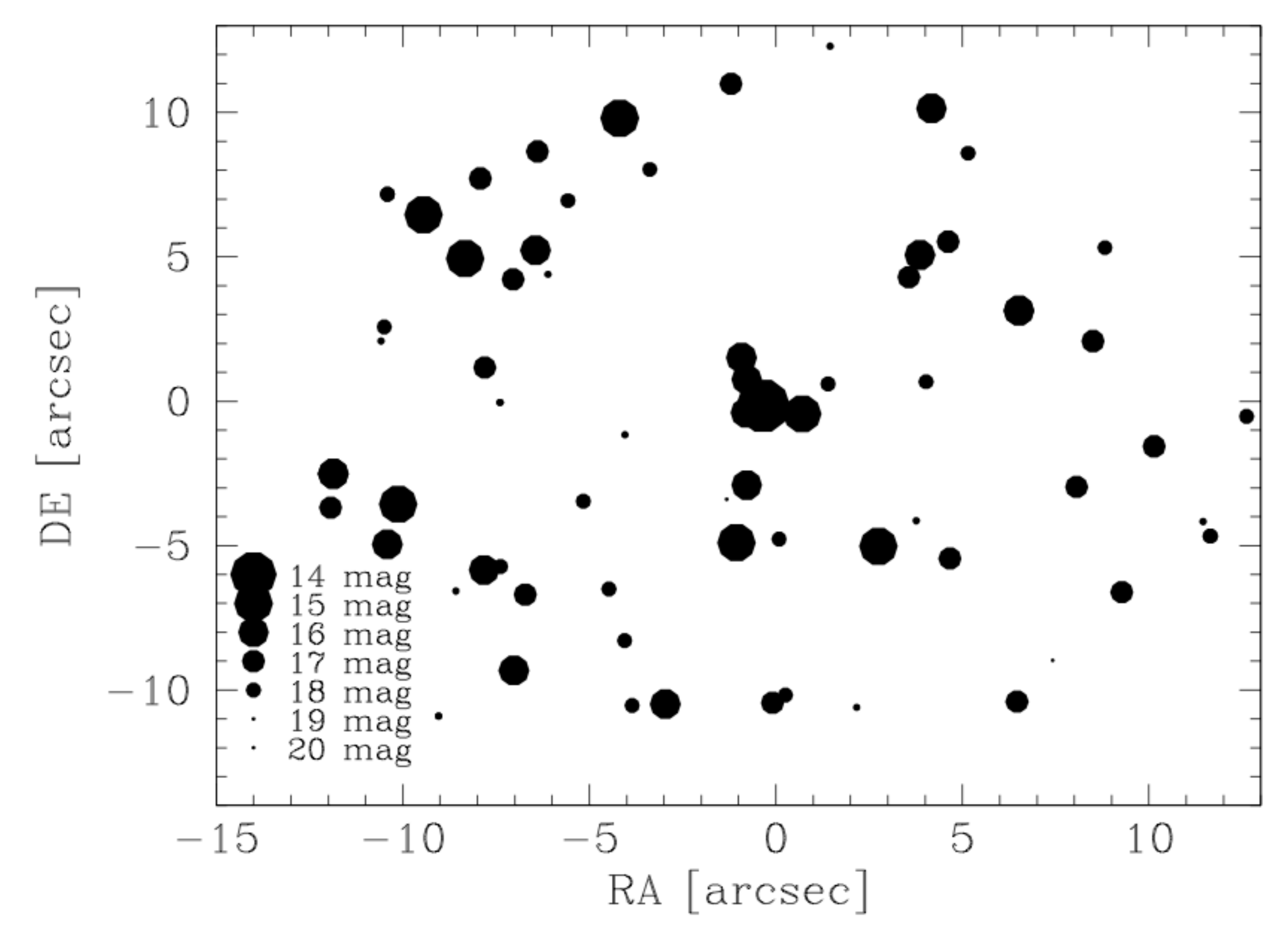}
\caption{Shown are the positions and brightnesses of all stars
detected by NaCo other than the targets in the anti-centre fields.
There is a notable excess of stars within 2\arcsec of the targets.}
  \label{BackgroundStars}
\end{figure}

Fig.\,\ref{BackgroundStars} shows the position and brightness of
  all stars, other than the targets that we detected in the
  anti-centre fields.  The sizes of the symbols indicate the
  brightness of the stars in the J-band.  The stars clearly cluster at
  the centre arround the targets. The probabilty that this
  distribution of stars in the field is just a chance coincidence is
  only $\sim 4\,10^{-6}$. This means that it is very unlikely that 
  so many stars are found within two arcsec of the targets hust by chance.
  Since background stars are expected to be homogenesously distributed
  over the field of view, the distribution of stars favours the
  hypothesis that they are physical companions.

\begin{figure}
\includegraphics[height=.25\textheight,angle=0]{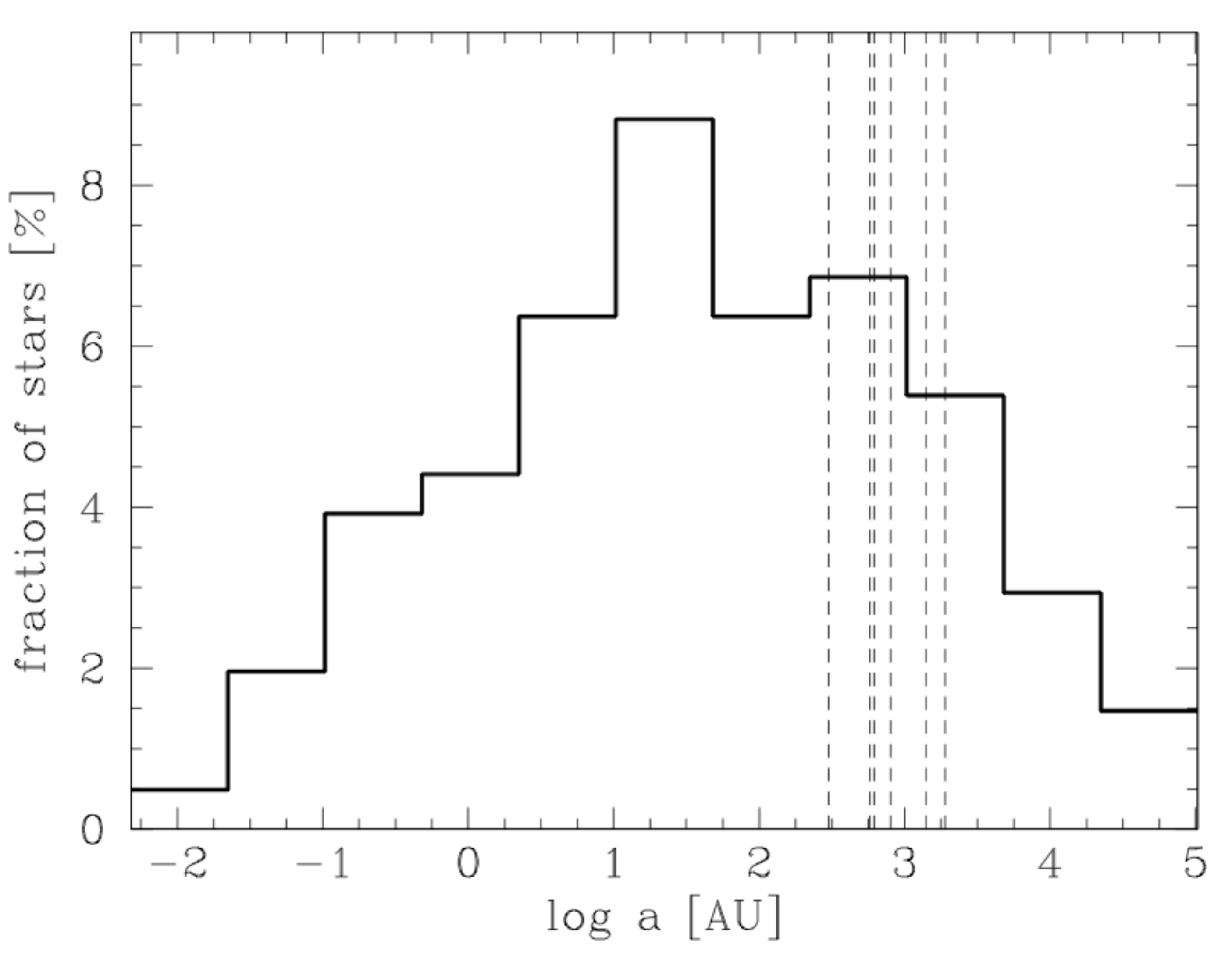}
\caption{Shown is the distribution of binaries in the solar 
neighbourhood taken from Duquennoy \& Mayor (\cite{duquennoy91}).
The dashed lines indicate the projected distances of the 
companion-candidates, assuming that they are physical companions.}
  \label{BinaryStars}
\end{figure}

Fig.\,\ref{BinaryStars} shows the distribution of binaries in the solar
neighbourhood from Duquennoy \& Mayor (\cite{duquennoy91}).  The
dashed lines are the projected distances of the CCs if we assume that
they are physical companions. Since the projected distances are close
to the maximum of the distribution of binaries, it is quite
plausible that they are physical binaries.

As discussed in Section~\ref{Sec:StarsCC}, we can calculate the
J-K colours for the assumption that the CCs are physical companions
($J-K_{phys}$), and for the assumption that they are unrelated
background stars ($J-K_{back}$).  $J-K_{phys}$, is derived by using the
brightness difference between the target and the CC to calculate the
spectral type of a physical companion, from which we obtain its $J-K$
colour, and $J-K_{back}$ from the average colour of stars within
10\arcmin of the target taken from 2MASS (Skrutskie et
al. \cite{2MASS}).  Since we obtained J- and K-images for three of the
CCs, we can compare the observed colour $J-K_{obs}$ with
$J-K_{phys}$) and $J-K_{back}$.

For the LRa01$\_$E1$\_$4719 CC, we derive $J-K_{obs}=0.4\pm0.3$,
$J-K_{phys}=0.50\pm0.07$ (G9V), and $J-K_{back}=0.64\pm0.36$
(Fig.\,\ref{4719phot}). The colour of the companion-candidate thus is
consistent with a physical companion, but this does not prove that it
is a companion because field stars have the same colour.  For
LRa02$\_$E1$\_$1715 CC we find $J-K_{obs}=1.1\pm0.2$, $J-K_{phys}=0.9\pm0.1$
(M4V), and $J-K_{back}=0.73\pm0.37$ (Fig.\,\ref{1715phot}). The
situation is the same as with LRa01$\_$E1$\_$4719 CC: the colour of the
companion-candidate is consistent with a physical companion but also
consistent with a background star.  The third object is the
companion-candidate of LRa02$\_$E2$\_$1136 (Fig.\,\ref{1136phot}) for which
we obtain $J-K_{obs}=0.9\pm0.2$, $J-K_{phys}=0.70\pm0.05$ (K4V-K5V),
and $J-K_{back}=0.73\pm0.37$.
 
Thus in all three cases, the colour of the companions is consistent
with physical companions as well as with an unrelated background star.
Table \ref{tab:CCoptical} thus gives the brightness in the optical
regime for both hypotheses. The distribution of stars within the
field of view makes it, however, more likely that they are physical
companions. This raises the question of whether the expected number of CCs
corresponds to the expected number of binaries with that
separation. As shown in Fig.\,\ref{BinaryStars}, we expect to find
only 6-7\% of the stars to be binaries with this separation but we found
of 28-35\% of the candiates have CCs.

One question that we cannot answer yet is whether the CCs are
eclipsing binaries by themselves and thus FPs.  Since 8\% of the
stars in the solar neighbourhood are triple stars (Tokovinin
\cite{tokovinin2008}) and since triple stars containing two eclipsing
late-type binaries are not that rare (e.g. Guenther et
al. \cite{guenther01}), it is possible that many CCs are FPs. An
alternative explanation is that planets form preferentially in binary
systems. More observations are thus needed to find out whether they
are physical companions or not and whether they are eclipsing binaries or
not.

\subsection{The rate of candidates, false positives and planets}

As mentioned in the introduction, most of the FPs are removed by the
detailed analysis of the light curves, by taking one image during
transit and one out of transit with a seeing-limited telescope, and by
spectroscopic observations.

Although the seeing-limited observations are not the subject of this
article, it is interesting to know what the total fraction of
candidates is that were identified as FPs and what the fraction of
stars with planets is amongst the candidates. In other words, how many
of the original candidates are FPs and how many stars have planets?
In IRa01 9872 stars were analysed, and 50 sources were classified as
planetary transit candidates, of which two are planet host stars
(Carpano et al \cite{carpano09}).  In LRa01 11408 stars were analysed,
51 sources were classified as planetary transit candidates, and four
stars hosting planets were found (Carone et al. \cite{carone12}).  In
LRc01 11408 stars were analysed, 42 sources were classified as
planetary transit candidates, and three planets and one brown
dwarf were found (Cabrera \cite{cabrera09}).  In SRc01, 6974 light
curves were analysed, and 51 candidates were found, but no planet has been
found yet (Erikson \cite{erikson12}).  In the case of LRa03, 5329 light
curves were analysed, and 19 candidates were found but no planet has been found
yet. For SRa03, 4169 light curves were analysed, 11 candidates were
identified, and three planets were found (Cavarroc \cite{cavarroc12}).  To sum it
up, 49160 stars were observed, 224 candidates were identified, and so far
12 planet host stars and brown dwarfs have been found in IRa01, LRa01,
LRa03, LRc01, SRc01, SRa03.  The frequency of planets amongst the
candidates in these fields thus is $\geq 5\%$.


In this work we studied objects in the field LRa01, LRa02, LRa03,
  LRa04, LRa06, LRa07, SRa01, SRa02, SRa03, SRa04, LRc02, and LRc07.
  In total CoRoT observed 88478 stars in these fields, 306 candidates
  were identified, and 18 planets have been found so far. Thus, 6\% of the
  candidates are planet host stars, which is quite similar to the
  results obtained for the fields for which detailed reports have been
  published. However, if we just take the top priority candidates,
  since we included only these ones in the follow-up observations with
  NaCO and CRIRES, and if we take only the fields LRa01 to LRa03 and
  SRa01 to SRa03, where the follow-up observations are almost
  completed, we find a somewhat different picture. In this case at
  least 21\% of the candidates are planet host stars. 
  We identiffied 15\% of the
  candidates as FPs using on/off photometry with
  seeing-limited telescopes, 13\% of the candidates
  because the spectral classification showed that the stars
  are either giants, or early-type stars or rotate too rapidly to allow
  precision radial-velocity measurements. We excluded another 17\% 
  using radial-velocity measurements, which showed that these objects
  are binaries. The NaCo/CRIRES observation removed another 9\% of
  the candidates from the list.  The remaining 25\% of the stars are
  simply too faint to carry out radial-velocity measurements, or the
  RV-amplitudes were too small to yield a detection.

\section{Conclusions}

Using adaptive optics imaging and high spectral resolution NIR
spectroscopy, we have investigated a sample of 25 CoRoT targets
for contamination of previous seeing-limited PSFs by FPs, i.e. very
close eclipsing binaries that would mimic the signature of a
transiting planet in the light curves obtained by CoRoT. Two of the
targets are in LRc fields where we {\it a priori} expected a high rate
of background sources. Only 13 of the 23 objects in the LRa field were
observed with NaCo and CRIRES. Of these six have CCs. Since for the
other ten objects we obtained either only CRIRES or only NaCo-data,
and since the CRIRES spectra often are not deep enough to exclude
all types of CCs, the true number of targets with CCs could even be
higher.  This relatively high rate of targets with CCs is, however,
roughly the same for Kepler.

\begin{acknowledgements}

We are grateful to the user support group of VLT for all their help
and assistance in preparing and carrying out the observations.  Some
of the data presented were acquired with the IAC80 telescope operated
at Teide Observatory of the Instituto de Astrof\'\i sica de Canarias.
This publication makes use of data products from the Two Micron All
Sky Survey, which is a joint project of the University of
Massachusetts and the Infrared Processing and Analysis
Center/California Institute of Technology, funded by the National
Aeronautics and Space Administration and the National Science
Foundation. This research has made use of the SIMBAD database,
operated at CDS, Strasbourg, France. The team at the IAC acknowledges
support by grants ESP2007-65480-C02-02 and AYA2010-20982-C02-02 of the
Spanish Ministerio de Ciencia e Innovaci\'on.  MONET (MOnitoring
NEtwork of Telescopes) is funded by the ``Astronomie \& Internet''
program of the Alfred Krupp von Bohlen und Halbach Foundation, Essen,
and operated by the Georg-August-Universit\"at G\"ottingen, 
McDonald Observatory of the University of Texas at Austin, and the
South African Astronomical Observatory. TRAPPIST is funded by the
Belgian Fund for Scientific Research (Fond National de la Recherche
Scientifique, FNRS) under the grant FRFC 2.5.594.09.F, with the
participation of the Swiss National Science Foundation (SNF).  We are
thankful to the Tautenburg observing team, particularly D. Sebastian,
M. Ammler-von Eiff, B. Stecklum, and Ch. H\"ogner, for helping us with
the NASMYTH observations.

\end{acknowledgements}

\appendix

\section{Detailed desciption of the stars with companion-candidates}
\label{Sec:StarsCCdetailed}

\subsection{LRa01$\_$E1$\_$2101}   

Seeing-limited images obtained with the 3.6m Canada France Hawaii
Telescope (CFHT; located at Mauna Kea, Hawaii, USA), the IAC 80 cm
(located at Iza\~na Tenerife, Spain), and the 1.2m robotic MONET
telescope (located at McDonald Observatory, USA) did not show any
stars in the field that became significantly fainter during
transit. The NaCo image shows a previously unrecognized star with
$J=16.3\pm0.1$ at a distance of $1\farcs76$ ($0\farcs92$ E and
$1\farcs51$ N) from the primary (Fig.\,\ref{NaCoImages}).  If it were an
unrelated star to the target and using the average colour of stars in
the field, we estimate that it would be about $V_{UCAC}\sim17.9$.  If
it were a physical companion, its brightness would correspond to that
of an M5V-star, which would be $V\sim 21.5$ and $R\sim19.7$. Thus, no
matter if the CC were related or unrelated to the primary, it is in both
cases bright enough to be an FP.  If the CC were a physical companion,
it would be at a distance of $\sim 300\,AU$ from the primary.

We obtained a spectrum with CRIRES of LRa01$\_$E1$\_$2101 in the range of
the CO bands (2284.1 to 2322.9 nm).  To search for additional
companions, we kept the CC outside of the slit.
Fig.\,\ref{L2101CRIRES} shows part of the CRIRES spectrum together
with a spectrum of a sunspot (Wallace \& Livingston
\cite{wallace92}).  The spectrum of LRa01$\_$E1$\_$2101 shows only the
CO lines of the primary but not of any other additional star.  Using
the cross-correlation function, we derived an upper limit for possible
additional companions. Fig.\,\ref{L2101cc} shows the cross-correlation
function of LRa01$\_$E1$\_$2101, together with that of a hypothetical M3V
companion star.  We can thus exclude that there is any additional CC
with a spectral type of M3V or earlier with a separation of
\asecdot{0}{3} or less.

\subsection{LRa01$\_$E1$\_$4719}   

Seeing-limited images obtained with the MONET telescope did not show
any contaminants. According to EXODAT, the spectral type of the object
is F8IV. However, in the NaCo images, the target is nicely resolved
into two stars with a separation of \asecdot{0}{8}.  This is shown in
Fig.\,\ref{NaCoImages} where the fainter component can be found
\asecdot{0}{71} W and \asecdot{0}{44} S of the brighter one.
To assess the nature of the companion, we obtained J- and K-band
images.  From the images, we derive $J=14.8\pm0.1$ and $K=14.5\pm 0.2$
for the primary and $J=15.8\pm0.1$, $K=15.4\pm 0.2$ for the secondary.

Fig.\,\ref{4719phot} shows the absolute brightness in the J-band
($M_J$), and the J-K colours, assuming that the two objects are at the
same distance, together with the absolute brightness and colours of
stars taken from L{\'e}pine et al. (\cite{lepine09}), Henry et
al. (\cite{henry2006}), and Bilir et al. (\cite{bilir09}).The
brightness as well as the $J-K$-colours of the secondary is in
agreement with being a physical G9V-companion at a projected distance
of $\sim 1900$ AU. If the CC were either bluer or redder, we would
know that it is an unrelated star.  Table \ref{tab:CCoptical} gives
the brightness of the companion in the optical regime for the case
that it is a physical companion, as well as for the case that it is an
unrelated background star.  In both cases the CC is bright enough to
be an FP.

The transit could thus either be on the primary, or on the
secondary. If it is on the primary, the size of the occulting object
would be $\sim 0.4$ $R_{Jup}$, and if it is on the secondary,
the size would
be $\sim 0.7$ $R_{Jup}$. Thus, in both cases the transiting object
could be a planet.

\subsection{LRa02$\_$E1$\_$1715}

The spectral type of this object is G2IV/V, as determined from a HIRES
spectrum taken with the Keck telescope.  Seeing-limited images with
CFHT, the 1.2m Leonard Euler telescope at ESO, La Silla, and the IAC
80cm telescope gave us the result that the transit is on target.
However, NaCo resolves the target into two components, which have a
separation of \asecdot{1}{52} (companion \asecdot{1}{40} W,
\asecdot{0}{60} N of the primary and $J=18.8\pm0.1$, $K=17.7\pm0.1$;
see Fig.\,\ref{NaCoImages}).  In order to constrain the nature of the
companion, we took J- and a K-band one. Fig.\,\ref{1715phot} shows
the absolute brightness in the J-band ($M_J$), assuming that the two
objects are at the same distance, together with the J-K colours of the
two stars. Both stars are slightly redder than the standard
stars. Since they are reddened by the same amount, the data are fully
consistent, with the hypothesis that their two stars are at the same
distance. Thus, the colour and the brightness difference between the
primary and secondary is consistent with an M4V star companion at a
projected distance of about 1400 AU.  If the transit were on the
primary, the occulting object would have $\sim 0.2$ $R_{Jup}$ and
where thus be in the planetary regime. In the case the transit is on the
secondary, the occulting object would have a radius between $\sim 1.5$
and $\sim 3$ $R_{Jup}$, corresponding to a low-mass star.

\subsection{LRa02$\_$E2$\_$1136}   

Images taken with the 1m ESA OGS telescope (Iza\~na Tenerife, Spain)
allow exclusion of FPs with $V<21$ at distances larger than three arcsec
from the target.  We obtained NaCo images in the J- and in the K-band
of this object. In both images the object is resolved into two stars
with a separation of \asecdot{0}{4} (primary: $J=12.6\pm0.1$,
$K=12.2\pm0.1$, secondary: \asecdot{0}{34} $\pm$ \asecdot{0}{03} E,
\asecdot{0}{16} $\pm$ \asecdot{0}{03} S, $J=14.5\pm0.1$,
$K=13.6\pm0.1$; Fig.\,\ref{NaCoImages}).  The primary is a
G0V star. Since not only the brightness difference but also the J-K
colour matches that of a K4V to K5V companion, it is likely that this
star is a physical companion at a projected distance of $\sim 800\,AU$
(Fig.\,\ref{1136phot}). Because the transit is only 0.3\% deep, it could
be either on the primary or the secondary. If it is on the primary,
the transiting object would be $\sim 0.6\,R_{Jup}$, and if it is on the
the secondary, it would be due to a $\sim 2.4\,R_{Jup}$ object. Thus it
could either be a planet or a low-mass star. Interestingly, the
NaCo images show another binary (Corot-ID 110676867,
2MASS06515971-0536425) in the same field of view, but the distance to
LRa02$\_$E2$\_$1136 is quite large, \asecdot{11}{44} (\asecdot{9}{56} E and
\asecdot{6}{56} N). The distance between the two stars is
\asecdot{0}{27}, and the two components are $J=15.5\pm0.2$ and
$J=15.6\pm0.2$.

\subsection{LRa03$\_$E2$\_$0861}


When making the acquisition image of LRa03$\_$E2$\_$0861 with CRIRES, we
recognized a CC. The CC is at a distance of \asecdot{1}{08} from the
primary (companion \asecdot{0}{78} E, \asecdot{0}{75} N of the
primary; Fig.\,\ref{NaCoImages}). A subsequent image taken with NaCo
showed that the CC is $J=16.4\pm0.1$. The star is thus bright enough
to be a FP (Table \ref{tab:CCoptical}).  If this star were a physical
companion, it would be an M4V star at a 600 AU.  Seeing-limited images
taken during transit with the Euler telescope and the CFHT telescope
show the transit, but the faint star detected with NaCo and CRIRES can
not be fully excluded as an FP. The situation is the same as for
LRa02$\_$E2$\_$1136: If the transit is on the primary, the object has the
size of a planet. If it is on the companion, it has the size of a star.
 
\subsection{LRa04$\_$E2$\_$0626} 


The NaCo image of this star shows a previously unrecognized star, which
is $J=16.8\pm0.1$, at a distance of \asecdot{0}{9} from the primary
(\asecdot{0}{8} E, \asecdot{0}{4} S; Fig.\,\ref{NaCoImages}).  If it
were a physical companion, it would be an M3V star at a 600 AU.
Images taken during transit with the 0.6m TRAPPIST telescope at ESO,
La Silla (Gillon et al. \cite{gillon11}), show part of the transit, but
the faint star detected with NaCo is not resolved and thus cannot be
fully excluded as an FP. The CRIRES spectrum does not show any
additional CC. Also for this object we have the same sitation as
above: The transiting object can have either the size of a planet or
a star. That depends on whether the transit is on the primary or the
secondary.


\subsection{LRc07$\_$E2$\_$0158} 

The NaCo image of this star shows a previously unrecognized star, which
is $J=14.6\pm0.1$, at a distance of \asecdot{0}{9} from the primary
(\asecdot{0}{0} E, \asecdot{0}{9} N;  Fig.\,\ref{NaCoImages}).  A deep image
taken with the EULER telescope, in fact, shows no sign of this star.
NaCo was thus required to detect it. A spectrum taken with the TLS-NASMYTH
spectrograph  mounted on the 2m Alfred-Jensch telescope at Tautenburg (Germany)
shows that this star has a spectral type F9IV or F9V.
If it were a physical companion, it would be an M1.5V star at 400 AU.
As shown in Table \ref{tab:CCoptical}, the star is bright enough to be
an FP.

\section{Detailed description of the stars without companion-candidates}
\label{Sec:StarsNoCCdetailed}

\subsection{LRa01$\_$E1$\_$0286} 

We observed LRa01$\_$E1$\_$0286 with NaCo and found six faint stars: The
closest star has a brightness of $J=15.7\pm0.1$ and a distance of
\asecdot{5}{0} (\asecdot{1}{0} E, \asecdot{4}{9} S). This star has
already been detected in the optical. The next closest is at a
distance of \asecdot{8}{6} (\asecdot{8}{0} W, \asecdot{3}{0} S) and is
only $J=17.4$. No additional star closer to the target was found.

We also obtained a CRIRES spectrum of this star. The spectrum shows
the CO lines with a strength corresponding to an M0V star that is
about 2-3 mag fainter. However, a TLS-NASMYTH spectrum shows 
that the star has a spectral-type G0V star. Since the strength of the CO lines
is what is expected for an early G star, the CO lines are thus
presumably the lines of the primary.

\subsection{LRa01$\_$E1$\_$2240} 

LRa01$\_$E1$\_$2240 was observed with NaCo but not with CRIRES. We found
three nearby stars that are at \asecdot{5}{6} (\asecdot{3}{6} W,
\asecdot{4}{3} N), \asecdot{6}{4} (\asecdot{3}{9} W, \asecdot{5}{1}
N), and \asecdot{7}{2} (\asecdot{4}{6} W, \asecdot{5}{5} N)
distance. The stars have a brightness of $J=17.8\pm0.1$,
$J=16.4\pm0.1$, and $J=17.9\pm0.1$, respectively.


\subsection{LRa01$\_$E1$\_$4667 } 

We obtained a spectrum with CRIRES of this star. This spectrum allows
exclusion of a companion star with a spectral type M3.5V or earlier.
Since CRIRES is also an AO-instrument, we can also use the aquisition
image to search for previously unrecognized CCs close to the
star. From the aquisition image taken in the K-band we can exclude a
companion of roughly the same brightness as the target within two
arcsec.

\subsection{LRa01$\_$E2$\_$0165, or CoRoT-7}

As part of this programme, we obtained deep images with NaCo and CRIRES
of the field surrounding the \object{CoRoT-7} object. The results are
discussed in detail in L{\'e}ger et al. (\cite{leger09}), where we
report that we did not find any CCs.  We do not discuss this object
further here but refer instead to the above-mentioned paper.

\subsection{LRa02$\_$E1$\_$1475} 

This star is an A4V star with a transit that is 0.3\% deep. This means
that the eclipsing object could have a size of about 0.1 $R_{sun}$,
corresponding to the size of Jupiter. From the depth of the transit
and the spectral type of the primary, we find that an FP would have to
be a star with a spectral type of K7V, or earlier that is being
eclipsed. We thus have to exclude stars with spectral type K7V 
to prove that the transit is on target. The aquisition image
shows the star 2MASS06512856-0348468 which has K=15.8 and thus 3.2 mag
fainter than the target in K. The images show three additional stars.
The first one is at a distance of \asecdot{2}{8}
south-east of the target, the second \asecdot{4}{8} (\asecdot{4}{3} W,
\asecdot{2}{3} N), and the third one at a distance of \asecdot{6}{5}
(\asecdot{6}{4} W, \asecdot{0}{7} S). All four stars had already been
observed with the CFHT and in all cases an FP was ruled out.  The
CRIRES aquistion image alone already allows physical
companions with spectral types earlier than K3V to be ruled out.  
Using the Mg 4383.23 $cm^{-1}$ , and Fe 4396.25 $cm^{-1}$ line, 
we can rule out companions
with spectral type F6V stars or earlier. The observation thus does not
fully rule out all possible FPs but still a large number of them.

\subsection{LRa02$\_$E1$\_$4601}   

For this object we carried out the most comprehensive study by taking
NaCo images in J and K and CRIRES spectra in both settings.  We can
thus use these data to assess which setting is the most
sensitive. 

The NaCo J-band image would have allowed us to detect a star of J=17
at a distance of \asecdot{0}{18} from the primary, a star of J=18 at
\asecdot{0}{23} distance, and a star with J=19 at \asecdot{0}{58}
distance.  The limits for the K-band image are \asecdot{0}{16} for a
star of $K=16$, \asecdot{0}{30}for K=17, and \asecdot{0}{51} for
$K=18$. If we assume that we want to detect a physical binary, than
the limit in both filters is almost the same, although the J-band
image is deeper. In both cases, we can exclude companions with
spectral types earlier than M2.5V for distances $\geq$ \asecdot{0}{2}
from the primary.  For distances $\geq$ \asecdot{0}{25}, the limit is an
M3.5V star and for distances $\geq$ \asecdot{0}{5} a M4.5V star.

\begin{figure}
\includegraphics[height=.22\textheight,angle=0]{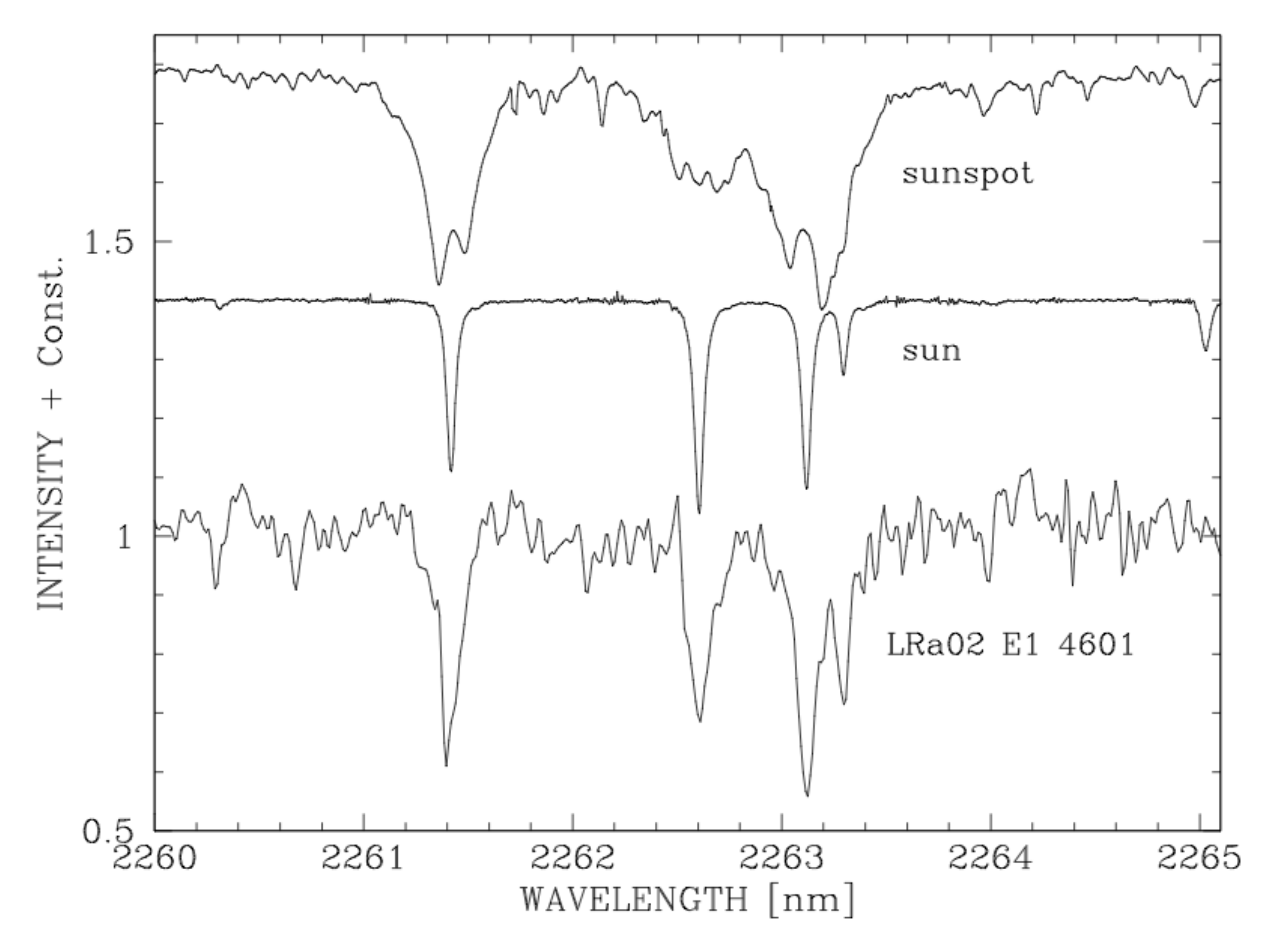}
\caption{Part of the CRIRES spectrum of LRa02$\_$E1$\_$4601 taken with the first
  setting, together with a spectrum of the sun and a sunspot. }
  \label{CRIRES-CoRoT24}
\end{figure}

\begin{figure}
\includegraphics[height=.22\textheight,angle=0]{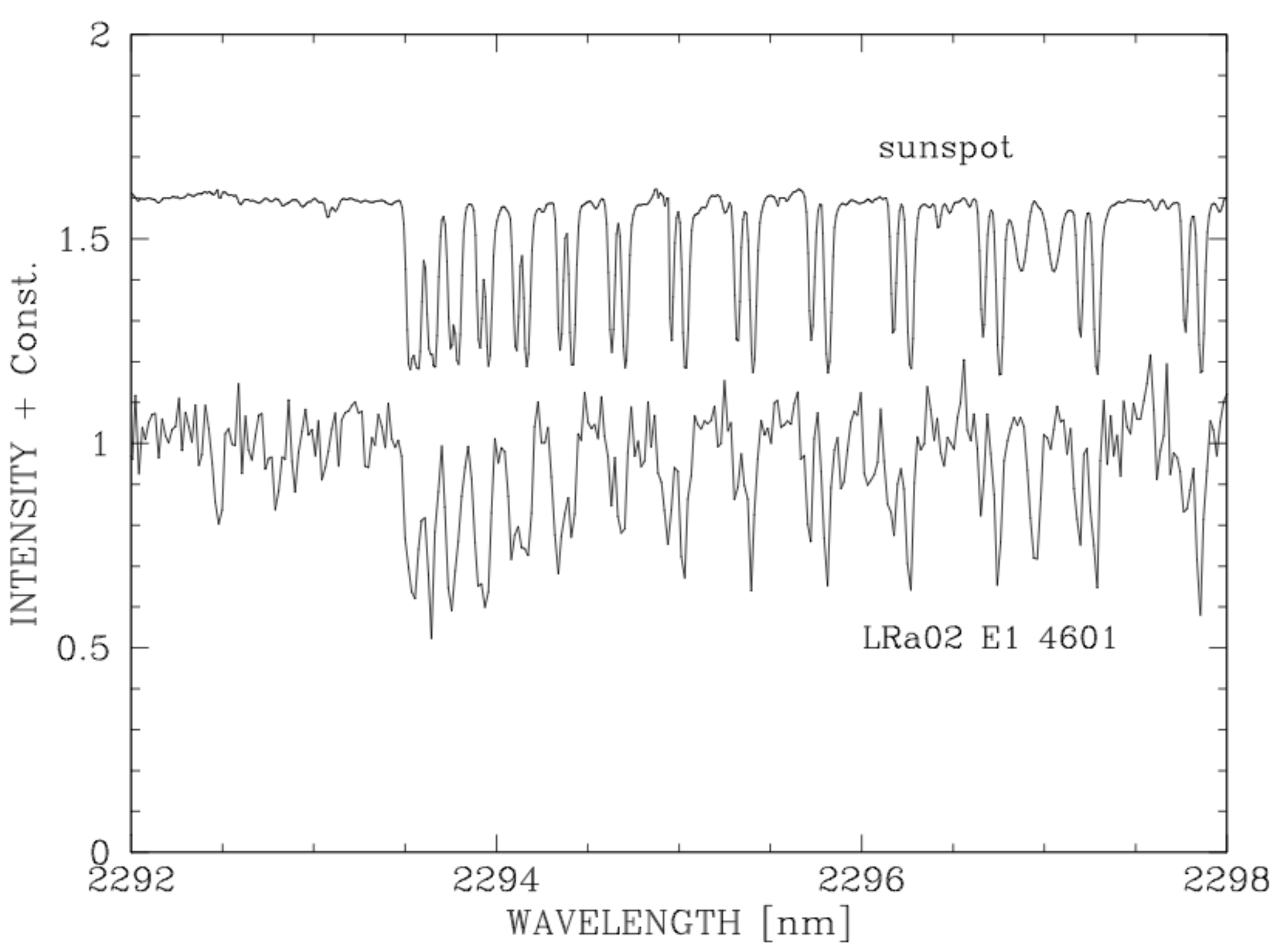}
\caption{Part of the CRIRES spectrum of LRa02$\_$E1$\_$4601 taken with the second
  setting, together with a spectrum of the sun and a sunspot.}
  \label{CRIRES-CoRoT24a}
\end{figure}

Fig.\,\ref{CRIRES-CoRoT24} shows part of the spectrum taken with the
first setting which contains prominent Ca\,I lines. This spectrum
allows exclusions of companion stars that are earlier than M1V.  
We note that
the Ca-lines at 2261.4 nm and 2263.1 nm (vacuum) are single in the
spectrum of the sun and LRa01$\_$E1$\_$2101 but double in the spectrum
of the sunspot due to the Zeeman effect (the magnetic field strength
of the spot is 3360 Gauss).  Fig.\,\ref{CRIRES-CoRoT24a} shows a part
of CRIRES spectrum taken with the second setting containing the
CO lines.  Although this spectral region is close to the edge of the
K-band and thus the sensitivity of CRIRES is lower, the number of
CO lines is so large that the sensitivity for detecting CCs is an
higher. Using this spectral range, we exclude companions earlier the
M2.5V. Thus, the second setting is more sensitive for detecting stars
with CO lines.

\begin{figure}
\includegraphics[height=.22\textheight,angle=0]{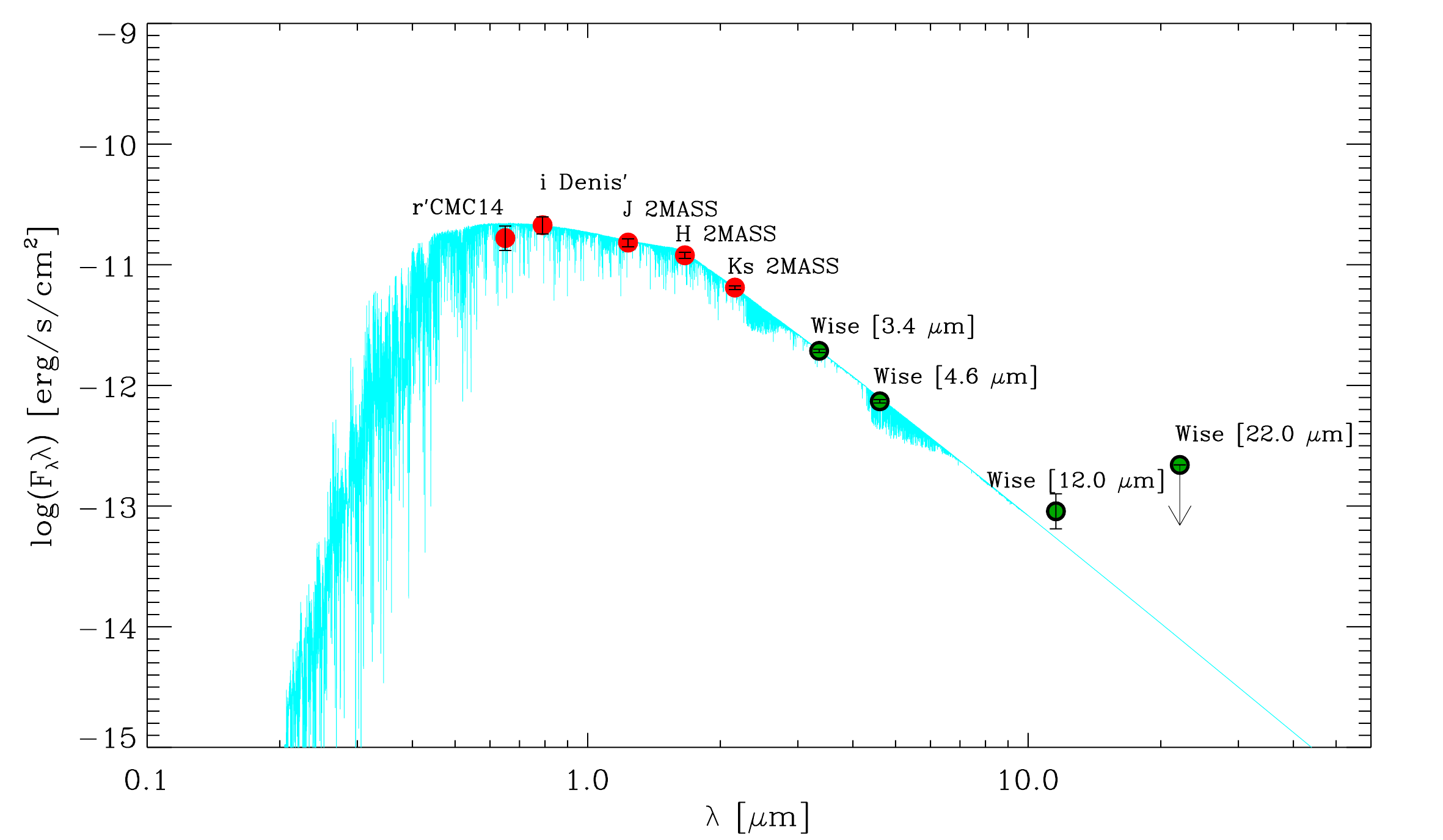}
\caption{Spectral energy distribution of LRa02$\_$E1$\_$4601. There is no
  excess in the IR. This is consistent with the NaCo and CRIRES
  results that the third object of this system is not a star.}
  \label{CRIRES-CoRoT24-SED}
\end{figure}

As a third approach we also derived the spectral energy distribution,
that shows no excess which would indicate a late-type companion
(Fig.\,\ref{CRIRES-CoRoT24-SED}). Altough this results conforms to the
previous ones, the SED method is significantly less sensitive than the
observations with NaCo and CRIRES.

\subsection{LRa02$\_$E2$\_$2057} 

The closest star to LRa02$\_$E2$\_$2057 seen in the NaCo image is
at a distance of \asecdot{5}{1} (\asecdot{2}{2} E,
\asecdot{2}{8} N), and it is $J=19.1\pm0.1$.  There is another star
with $J=21.1\pm0.1$ at a distance of \asecdot{7}{4} (\asecdot{7}{4} E,
\asecdot{0}{0} N).  Since an FP could be as faint as $J=20.9\pm1.0$,
both stars are bright enough to be FPs. Observations with the IAC 80
cm telescope and the CFHT show that neither star is an FP.  The CRIRES
spectrum of LRa02$\_$E2$\_$2057 excludes companions with spectral
types M0V or earlier.  This means that the CRIRES spectrum allows
companions with $R\leq18.6\pm0.2$ to be excluded. However, an FP could be as
faint as $R=22.5$ (corresponding to a companion with spectral-type
M3V).  There is thus still the possibility for an FP but this companion
would have to be much closer than one arcsec to the target and it
would have to have a spectral type in the range between M1V and
M3.5V. A companion with exactly these properties is not
very likely.

\subsection{LRa02$\_$E2$\_$3804} 

The observations of LRa02$\_$E2$\_$3804 with NaCo did not show any CC. The
closest star to LRa02$\_$E2$\_$3804 is already at a distance of
\asecdot{10}{0} (\asecdot{5}{2} W, \asecdot{8}{6} N), and it is
$J=18.3\pm0.1$.


\subsection{LRa03$\_$E2$\_$0678} 

The expected brightness of an FP is $J\leq18.9\pm1.0$.  The closest
star found is at a distance of \asecdot{9}{8} (\asecdot{2}{1} E,
\asecdot{9}{6} N) and has $J=19.4\pm0.1$.  Observation with
the CFHT and with the IAC 80cm telescope rules out that this star is an
FP. The CRIRES observations exclude companions earlier than M1V,
corresponding to a star with $R=13.9$. Also in this case, the CRIRES
observations are not deep enough to exclude close companions of 
very-late spectral type, but they exclude at least all companions earlier
than M1V.

\subsection{LRa03$\_$E2$\_$1326} 

In the case of LRa03$\_$E2$\_$1326, the closest stars found in the NaCo
images are already at distances of \asecdot{8}{3} (\asecdot{6}{4} E,
\asecdot{5}{2} N) and\asecdot{10}{5} (\asecdot{0}{1} E,
\asecdot{10}{4} S) and have a brightness of $J=16.9\pm0.1$ and
$J=17.9\pm0.1$.


\subsection{LRa06$\_$E2$\_$5287}

The NaCo image obtained of LRa06$\_$E2$\_$5287 shows ten stars, 
in addition 
to the target. The closest one is at a distance of \asecdot{3}{6}
(\asecdot{1}{3} E, \asecdot{3}{4} S) but it is J=20.3. The next
closest star is at a distance of \asecdot{5}{0} (\asecdot{3}{7} W,
\asecdot{3}{4} N), but it is even J=21.4. These stars should thus are too
faint to be FPs.

CRIRES spectrum was taken that shows weak CO lines. A subsequent spectrum
taken with the TLS-NASMYTH spectrograph shows that the primary is a
G0V star. Since such a star has weak CO lines, we interpret the lines
as coming from the target itself.

\subsection{LRa07$\_$E2$\_$3354} 

Like LRa02$\_$E1$\_$1475, LRa07$\_$E2$\_$3354 an early-type
star.  The aquisition image taken with CRIRES shows no star within
\asecdot{10}{0} arsec of the target. For the CRIRES observation, we
used the first setting and the Mg 4383.23 $cm^{-1}$ , and Fe 4396.25
$cm^{-1}$ to exclude companions. Given that a TLS-NASMYTH spectrum
shows that it is a B9 star, we can rule out companions earlier than
A6V. Since the transit is 3.9\% deep the eclipsing object could still
well be a late K or early M star.

\subsection{LRc02$\_$E1$\_$0591}

\begin{figure}
\includegraphics[height=.25\textheight,angle=0]{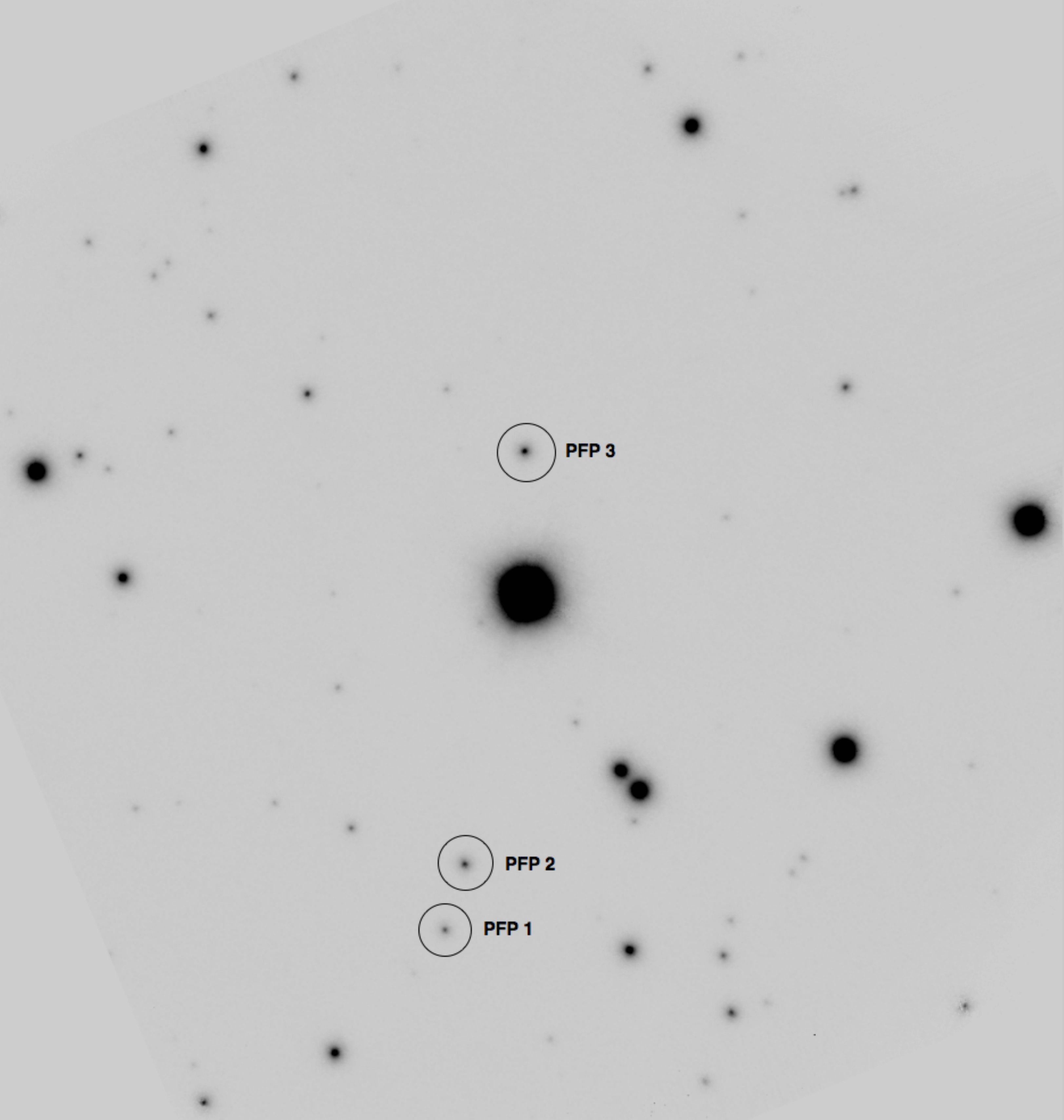}
\caption{NaCo image of LRc02$\_$E1$\_$0591 taken in the J-band. Marked as
PFP 1,2,3 are stars that became fainter during the transit (``potential 
FP''). See also Table \ref{tab:0591}. North is up and east is left.}
\label{image:0591}
\end{figure}

This object is is located in the  "galactic center" eye
(LRc-fields) of CoRoT.  The density of stars in this region is much
higher than in the  "galactic anti-center" eye (LRa-fields).
The large number of stars in this field made it necessary to obtain
two images with NaCo.  One was taken during transit and the other
out of transit.  Fig.\,\ref{image:0591} shows the image taken out of
transit. Table \ref{tab:0591} gives the brightness measurements and
their differences obtained in the two images. Three stars, labeled
PVP1, PVP2, and PVP3 in Fig.\,\ref{image:0591}, and Table
\ref{tab:0591}, were a little bit fainter during transit than out of
transit.  Since potential FPs of LRc02$\_$E1$\_$0591 have to be
$J\leq17.6\pm1.6$ (see Table \ref{tab:NaCo}), these stars could
potentially be FPs. However, images taken with the IAC 80cm and the
Euler telescope show that all of them are too faint in the optical and
thus can not be FPs.

\begin{table*}
  \caption{On-off photometry of stars in the field of LRc02$\_$E1$\_$0591}
\setlength{\tabcolsep}{3pt} 
 \begin{tabular}{ccccrl}
\noalign{\smallskip}
 \hline 
\noalign{\smallskip}
RA    & DEC     & out-of-transit & in-transit & $\Delta$ J& remarks \\
h:m:s & d:m:s &  J [mag]       & J [mag]    & [mag]    & \\
\noalign{\smallskip}
 \hline  
\noalign{\smallskip} 
$18^h 42^m 39\fs78$ &  $06\degr 12\arcmin58\farcs9$  &  $18.3\pm0.2$  &  $18.2\pm0.2$  &    $0.1\pm0.3$  & \\
$18^h 42^m 39\fs61$ &  $06\degr 13\arcmin05\farcs2$  &  $15.3\pm0.2$  &  $14.9\pm0.2$  &    $0.4\pm0.3$  & 2MASS18423960+0613053, J=$15.365\pm0.166$\\
$18^h 42^m 39\fs61$ &  $06\degr 13\arcmin13\farcs8$  &  $18.5\pm0.2$  &  $18.3\pm0.2$  &    $0.2\pm0.3$  & \\
$18^h 42^m 39\fs85$ &  $06\degr 13\arcmin20\farcs1$  &  $16.4\pm0.2$  &  $16.2\pm0.2$  &    $0.2\pm0.3$  & 2MASS18423986+0613198, J=$16.433\pm0.115$ \\
$18^h 42^m 39\fs95$ &  $06\degr 13\arcmin04\farcs5$  &  $15.2\pm0.2$  &  $15.1\pm0.2$  &    $0.1\pm0.3$  & \\
$18^h 42^m 39\fs95$ &  $06\degr 13\arcmin00\farcs4$  &  $17.3\pm0.2$  &  $17.0\pm0.2$  &    $0.3\pm0.3$  & \\
$18^h 42^m 40\fs11$ &  $06\degr 13\arcmin08\farcs9$  &  $12.6\pm0.2$  &  $12.1\pm0.2$  &    $0.5\pm0.3$  & 2MASS18424010+0613088, J=$12.414\pm0.027$ \\
$18^h 42^m 40\fs11$ &  $06\degr 13\arcmin12\farcs3$  &  $18.1\pm0.2$  &  $19.0\pm0.2$  &    $-0.9\pm0.3$  &  PFP1$^1$ \\
$18^h 42^m 40\fs21$ &  $06\degr 13\arcmin02\farcs5$  &  $18.4\pm0.2$  &  $18.5\pm0.2$  &    $-0.1\pm0.3$  &  \\
$18^h 42^m 40\fs23$ &  $06\degr 13\arcmin13\farcs8$  &  $20.4\pm0.2$  &  $20.6\pm0.2$  &    $-0.2\pm0.3$  & \\
$18^h 42^m 40\fs24$ &  $06\degr 13\arcmin00\farcs9$  &  $19.0\pm0.2$ &  $19.5\pm0.2$  &     $-0.5\pm0.3$  & PFP2$^1$ \\
$18^h 42^m 40\fs45$ &  $06\degr 13\arcmin13\farcs7$  &  $18.4\pm0.2$  &  $18.4\pm0.2$  &    $0.0\pm0.3$  & \\
$18^h 42^m 40\fs61$ &  $06\degr 13\arcmin15\farcs5$  &  $19.2\pm0.2$  &  $19.8\pm0.2$  &    $-0.6\pm0.3$  & PFP3$^1$ \\
$18^h 42^m 40\fs67$ &  $06\degr 13\arcmin12\farcs8$  &  $20.0\pm0.2$  &  $20.4\pm0.2$  &    $-0.4\pm0.3$  & \\
$18^h 42^m 40\fs75$ &  $06\degr 13\arcmin09\farcs3$  &  $17.2\pm0.2$  &  $16.9\pm0.2$  &    $0.3\pm0.3$  & \\
$18^h 42^m 40\fs88$ &  $06\degr 13\arcmin11\farcs8$  &  $15.9\pm0.2$  &  $15.7\pm0.2$  &    $0.2\pm0.3$  & 2MASS18424088+0613117,  J=$15.836\pm0.090$ \\
\noalign{\smallskip}
 \hline 
\noalign{\smallskip}
 \end{tabular}
 \label{tab:0591}
\\
$^1$ Star became significantly fainter during transit (``potential FP''), see also Fig.\,\ref{image:0591}.\\
\end{table*}

\subsection{SRa01$\_$E1$\_$0770} 

The two closest stars to SRa01$\_$E1$\_$0770 are at a distance of
\asecdot{6}{2} (\asecdot{5}{2} E, \asecdot{3}{5} S) and
\asecdot{11}{0} (\asecdot{1}{2} E, \asecdot{10}{9} N) and have
$J=18.8\pm0.1$, and $J=17.8\pm0.1$.  Since we have to remove FPs with
$J\leq17.5\pm1.2$, the second star is just bright enough that it could
be an FP. However, seeing-limited observations are sifficient to find out
if this star is an FP or not. The situation with the CRIRES
observations is the same as for the other stars. The spectra allow 
exclusion of companions earlier than M0V, corresponding to stars of
$R=17.0$, but this is not sufficient to fully exclude all types of FPs.

\subsection{SRa02$\_$E1$\_$1011} 

Using a spectrum obtained with the TLS-NASMYTH spectrograph,
we derived the
spectral type of this star, which is F6V. The NaCo image shows two
additional stars.  They are $J=18.9\pm0.1$ and $J=16.2\pm0.1$ and
are at distances of \asecdot{8}{9} (\asecdot{5}{6} E, \asecdot{6}{9} N),
and \asecdot{9}{8} (\asecdot{7}{8} E, \asecdot{5}{8} S), respectively. 
Both stars are thus well separated from the primary, and observations with
seeing-limited telescopes could show if these are FPs or not. The
CRIRES observations permit us to exclude companions earlier than M0V
corresponding to stars with $R\geq17.7$.  As before, the CRIRES
observations are not deep enough to exclude companions of very
late-type, but they do allow the exclusion of most of them.

\subsection{SRa03$\_$E2$\_$1073} 

A TLS-NASMYTH spectrum shows that this star has a spectral type F3V.
The CRIRES spectrum allows exclusions of a companion that is 1.7 mag
fainter in K and has CO lines like an M0V star. However, this
does not expclude stars with spectral type K2V because the CO lines
in these stars are weaker.  Since the CO lines become progressively
shallower for earlier type stars, an eclipsing binary with spectral
type G cannot be fully excluded.

\subsection{SRa03$\_$E2$\_$2355} 

The three closest stars detected in the NaCo images of SRa03$\_$E2$\_$2355
are $J=17.0\pm0.1$ (distance of \asecdot{3}{0}; \asecdot{0}{8} E,
\asecdot{25}{2} S), $J=20.2\pm0.1$ (distance of \asecdot{4}{2};
\asecdot{4}{0} E, \asecdot{1}{2} S), and $J=18.1\pm0.1$ (distance of
\asecdot{9}{2}; \asecdot{4}{0} E, \asecdot{8}{3} S).


\subsection{SRa04$\_$E2$\_$0106, CoRoT-32}

This object is CoRoT-32. Since it is discussed in Gandolfi
et al.  (\cite{gandolfi2013}), we will just briefly mention the results
here.  We obtained a spectrum and an aquisition image with
CRIRES. The limiting magnitude of the aquistion image is $K \geq 13.9$.
No additional stars were detected within  \asecdot{12}{4} of the target, which
means that physical companions with a spectral type M2V or earlier can
be excluded. The CRIRES spectrum itself allows exclusion of
physical companions with a spectral type M3V or later.

\end{document}